\PassOptionsToPackage{protrusion, expansion}{microtype} 

\documentclass[sigconf,screen]{acmart}

\usepackage{amsmath, amsthm}
    
\usepackage{graphicx, xcolor} 
\usepackage{hyperref, url}
\usepackage{algorithmic}
\usepackage{algorithm}
\usepackage{comment}
\usepackage{booktabs, multirow}
\usepackage{caption}
\usepackage[tight,footnotesize]{subfigure}
\usepackage{enumitem}
\usepackage{lipsum}
\usepackage{setspace}
\usepackage{diagbox}
\usepackage{soul}
\usepackage{textcomp}
\usepackage{xspace}
\usepackage{pifont}
\usepackage{subcaption}

\usepackage{tikz}

\usepackage{footnote}
\makesavenoteenv{tabular}
\makesavenoteenv{table}

\usepackage{graphicx}

\usepackage[acronym]{glossaries}
\makeglossaries
 
\newglossaryentry{fclayer}{
    name={\texttt{fc}-layer},
    description={fully connected layer}
}
 
\newglossaryentry{szfull}{
    name=\textsc{SZ lossy compression},
    description={SZ lossy compression}
}

\newglossaryentry{deepsz}{
    name=\textsc{DeepSZ},
    description={DeepSZ}
}

\newglossaryentry{dataArray}{
    name=\textsc{data array},
    description={}
}

\newglossaryentry{indexArray}{
    name=\textsc{index array},
    description={}
}
 
\newacronym{gcd}{GCD}{Greatest Common Divisor}
\newacronym{lcm}{LCM}{Least Common Multiple}

\graphicspath{{./figures/}}
\DeclareGraphicsExtensions{.pdf}

\newcommand{\tech}{\textsc{LLMFI}\xspace}

\begin{document}
\title{Not All Errors Are Equal: A Systematic Study of Error Propagation in Large Language Model Inference}

\author{Yafan Huang}
\affiliation{%
  \institution{University of Iowa}
  \city{Iowa City}
  \state{IA}
  \country{USA}
}
\email{yafan-huang@uiowa.edu}

\author{Sheng Di}
\affiliation{%
  \institution{Argonne National Laboratory}
  \city{Lemont}
  \state{IL}
  \country{USA}
}
\email{sdi1@anl.gov}

\author{Guanpeng Li}
\affiliation{%
  \institution{University of Florida}
  \city{Gainesville}
  \state{FL}
  \country{USA}
}
\email{liguanpeng@ufl.edu}

\begin{abstract}
Large language models (LLMs) are increasingly integrated into high-performance computing (HPC) workflows, accelerating scientific discovery through diverse perspectives such as code generation and domain-specific decision-making. Yet, how soft errors propagate and affect LLM inference remains largely unexplored. To bridge this gap, we present a comprehensive study on error propagation in LLM inference, enabled by our proposed \tech, a configurable and deterministic fault-injection framework. Using \tech, we systematically inject faults across three open-weighted LLMs and thirteen representative tasks, covering reasoning, multilingual, mathematical, and coding domains. In addition, we conduct fine-grained case studies that reveal critical vulnerability patterns. Overall, our study yields 17 takeaways that advance the understanding of error propagation in LLM inference and introduces four low-overhead directions to improve reliability through software-only modification, offering practical guidance for future error detection and mitigation\footnote{\textit{Accepted at the 2026 ACM International Conference on Supercomputing (ICS 2026).}}.
\end{abstract}

\keywords{Fault Tolerance, Large Language Model Inference, Error Mitigation}

\maketitle

\setlength{\textfloatsep}{6pt}

\section{Introduction}
As modern high-performance computing (HPC) systems continue to scale, soft errors, caused by high-energy particles or silicon decay, have become increasingly more prevalent~\cite{tiwari2015understanding,nie2016large,papadimitriou2021demystifying}.
Such transient faults manifest as unexpected bit flips in hardware components, silently propagating through program execution and altering outcomes without visible warning~\cite{li2018modeling,dixit2021silent,sun2025demystifying}.
Their silent yet destructive nature makes them notoriously difficult to detect or mitigate, posing a major threat to HPC.
Indeed, resilience has been identified as one of the top 10 challenges for exascale systems~\cite{lucas2014top}.

Meanwhile, large language models (LLMs) are increasingly being integrated into HPC workflows to assist scientific discovery~\cite{ding2023hpc,yin2024chathpc,nichols2024hpc}, allowing applications from parallel code generation~\cite{nichols2024can} to domain-specific decision making~\cite{li2024cllmate,wang2024exploring}.
For instance, recent work~\cite{li2024cllmate} demonstrates that, given only meteorological observations, LLM inference can predict extreme weather types and affected regions.
In such cases, a single bit-flip can corrupt intermediate computations, spread across the inference process, and ultimately change the LLM predictions, leading to wrong conclusions~\cite{cavelan2019detection}.

To ensure that scientists can make reliable discoveries, it is essential to understand how errors propagate during LLM inference.
Studying this enables analysis without expensive profiling, reveals which components and operations are most vulnerable, and hence provides valuable insight for designing error mitigation mechanisms.
In the past, error propagation has been extensively studied for traditional HPC programs~\cite{li2018modeling,anwer2020gpu,yang2021sugar,laguna2016ipas,mahmoud2018optimizing}, such as matrix multiplications, where computation is relatively more structured.
In comparison, understanding error propagation in LLM inference is fundamentally different and poses three critical challenges.

\begin{figure}[t]
\centering
\includegraphics[width=1.0\columnwidth]{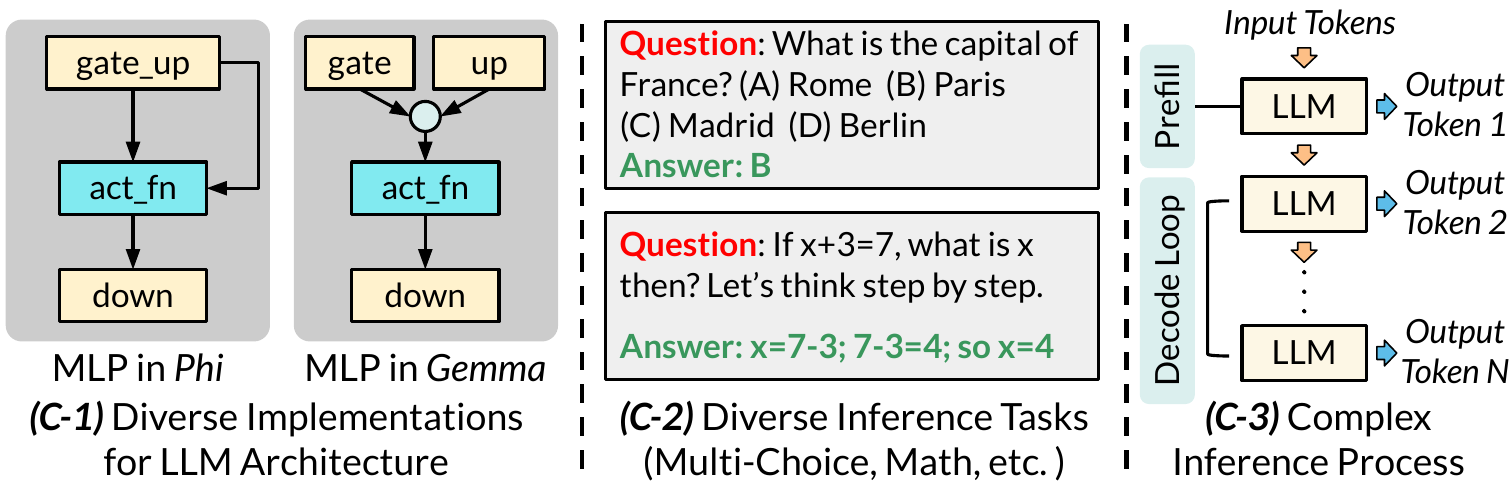}
\vspace{-3mm} 
\caption{Challenges for modeling errors for LLM inference.}
\label{fig:challenges-in-llm-inference}
\end{figure}

\textit{(C-1) Diverse Implementations for LLM Architecture.} 
Although state-of-the-art LLM models share a Transformer-based paradigm~\cite{vaswani2017attention}, their implementations vary substantially across leading AI companies~\cite{Abdin2024Phi3TR,team2024gemma,liu2024deepseek}.
Even seemingly simple components such as the multi-layer perceptron (MLP) differ in structure and parameterization (Figure~\ref{fig:challenges-in-llm-inference}, left), not to mention the more complex self-attention mechanisms and entire forward pipelines.
Modeling soft errors requires fault injection at diverse locations, yet these locations differ significantly among models with billions of parameters.

\textit{(C-2) Diverse Inference Tasks.}
LLMs are designed to handle a wide range of tasks, from reasoning-oriented benchmarks such as multiple-choice question answering~\cite{clark2019boolq,clark2018think} to generative workloads like mathematical problem solving~\cite{cobbe2021training,lightman2023let} (Figure~\ref{fig:challenges-in-llm-inference}, middle).
Different tasks exhibit distinct resilience behaviors: perturbations that are catastrophic for one task may be masked or corrected in another.
This task-dependent resilience complicates error modeling and makes generalization across workloads non-trivial.

\textit{(C-3) Complex Inference Process.}
As illustrated in Figure~\ref{fig:challenges-in-llm-inference} (right), LLM inference is an autoregressive process with multiple stages.
While prefill encodes the initial context and produces the first token, the subsequent decoding loop incrementally extends the output token sequence while updating the context.
Understanding how errors propagate across these coupled stages requires in-depth insight into the model’s runtime behavior, where the inherent non-determinism of LLM inference further exacerbates this challenge.

To address these challenges, we design \tech, a fault injector that enables, for the first time, a fine-grained modeling of error propagation for LLM inference.
Unlike existing fault injection frameworks that are confined to fixed LLM architectures or overlook the unique characteristics of inference stages~\cite{mahmoud2020pytorchfi,sun2025demystifying,dai2025ft}, \tech enables flexible and reproducible fault injection across LLMs, tasks, internal submodules, and inference stages.
Using \tech, we perform a comprehensive fault injection study using open-weighted LLMs from leading AI organizations and various representative tasks, injecting multiple fault patterns at different inference stages.
Insights drawn from these large-scale experiments further guide a series of fine-grained analyses, uncovering how errors propagate under various conditions.
From these, we distill 17 takeaways and 4 mitigation directions that not only redefine how reliability should be analyzed in LLM inference but also provide actionable guidance in practice for LLM resiliency, enabling developers to \textit{locate} where and when faults occur, \textit{improve} reliability through lightweight adjustments, and \textit{mitigate} errors through targeted protection once vulnerable components are identified.
\textit{To the best of our knowledge, this is the first work to propose a fault injector for detailed LLM inference stages and to perform a systematic study of error propagation. Our insights enable the design of concrete mitigation prototypes, bridging empirical characterization and practical reliability enhancement.}

The main contributions are summarized as follows:
\begin{itemize}
    \item We design \tech, a configurable fault injector, which is task-agnostic, model-agnostic, and inference-stage-aware. It supports deterministic execution and diverse fault models, including computation and memory faults with controllable bit-flip patterns, enabling systematic modeling of error propagation in LLM inference.
    \item We evaluate across 3 open-weighted LLMs and 13 representative tasks, covering reasoning, multilingual understanding, math solving, and code generation. For each model-task pair, we fine-tune prompts and report raw accuracy under various fault configurations.
    \item Based on the observations, we perform 6 case studies to analyze error propagation under various perspectives, such as layer- and stage-wise fault injection. These case studies, along with the aforementioned observations, generate 17 takeaways that characterize LLM resilience behavior.
    \item Inspired by these takeaways, we introduce 4 practical solutions to mitigate soft errors in LLM inference, providing insights for future research on LLM fault tolerance.
\end{itemize}

We summarize 17 takeaways and insights briefly as follows:
\begin{itemize}
\item \textit{Global and stage-level sensitivity}. Error resilience depends on when and where a fault occurs. Prefill and first-token stages are more vulnerable, suggesting that targeted protection at critical stages can yield maximal benefit.
\item \textit{Architectural and layer-wise dynamics}. Internal structures matter: dimension expansion mitigates, whereas reduction amplifies, error propagation. Early Transformer blocks and MoE architectures exhibit self-correction, indicating opportunities for resilience-aware architectural design.
\item \textit{Numerical and representation-level effects}. Numeric encoding directly impacts resilience: For example, specific exponent bits dominate corruption severity using \texttt{\small float16}, while formats exhibit varying recovery behaviors. 
\end{itemize}

The 4 proposed directions for mitigating error propagation are:

\begin{itemize}
    \item \textit{Mixed-precision Inference}: Employing different data
    types for specific components of LLM inference to enhance
    error resilience while preserving accuracy.
    \item \textit{Selective Checksum-based Protection}: Selectively deploying checksum-based protection only to layers that are dimension-preserving or dimension-reducing.
    \item \textit{Dual First-token Generation}: Reusing language context, generating the first output token twice, and comparing two copies before entering the decode loop.
    \item \textit{Dynamic Shot Inference}: Adaptively extending the conversation context length by increasing the number of shots to improve error resilience.
\end{itemize}

\section{Background}

In this section, we first break down the process of LLM inference and then introduce related works that motivate this study.

\subsection{Large Language Model Inference}

Existing research for LLM systems can be broadly categorized into two stages: training~\cite{zhao2024galore,yao2025holmes,bang2024vtrain,yu2025exploring} and inference~\cite{patel2024splitwise,stojkovic2025tapas,lazuka2024llm}.
Training is an offline process routinely conducted on massive GPU clusters; LLM inference, by contrast, is an online process that generates answers in response to user prompts.
Also, inference requires far fewer resources -- a single NVIDIA A100 GPU is sufficient to serve interactions with a $\sim$20B-parameter model.
However, because inference is invoked at extremely high frequency across applications, its cumulative computational cost exceeds that of training~\cite{patel2024splitwise,erdil2025inference}.

\begin{figure}[ht]
\centering
\includegraphics[width=1.0\columnwidth]{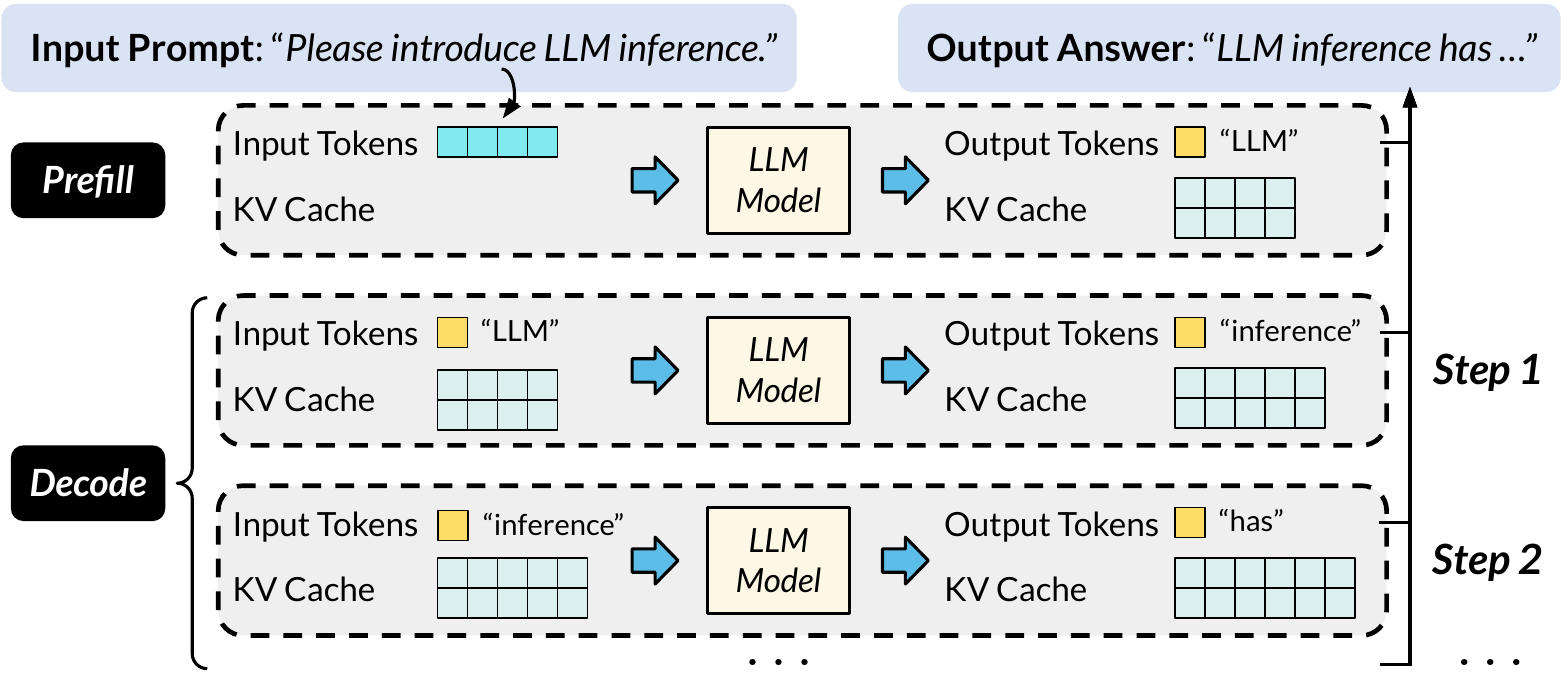}
\vspace{-6mm} 
\caption{Illustration of \textit{Prefill-Decode} LLM inference process.}
\label{fig:llm-inference}
\end{figure}

Figure~\ref{fig:llm-inference} illustrates the LLM inference process.
After a user provides an input prompt, the CPU tokenizes the text into discrete tokens, which are then passed to  the GPU for LLM model execution.
Modern Transformer-based LLMs follow a \textit{Prefill-Decode} inference paradigm~\cite{wolf2019huggingface} that leverages the GPU's massive parallelism.
In the \textit{Prefill} stage, LLM processes all input tokens, generates the first output token, and constructs the key-value (KV) cache.
The KV cache, stored in GPU memory, encodes the conversation context and is reused in every subsequent forward pass.
The \textit{Decode} stage then proceeds iteratively: at the $i$-th step, the model consumes the $i$-th generated token together with the preserved KV cache, produces the $(i+1)$-th token, and updates the cache with the new context.
The first decode step is particularly important, as its input token is generated from the prefill stage and therefore strongly influences the overall answer~\cite{zhong2024distserve}.
After decoding, the CPU converts the generated tokens back into natural language through de-tokenization, yielding the final human-readable response.
The computational workload for the whole LLM inference process is dominated by large-scale matrix multiplications executed on the GPU, whereas CPU-side (de-)tokenization incurs negligible overhead~\cite{hong2024flashdecoding++}.


\subsection{Related Works and Motivation}

With shrinking transistor dimensions, soft errors have become prevalent in modern large-scale HPC systems~\cite{tiwari2015understanding}. 
These errors manifest as bit flips in hardware and can silently corrupt program outputs without notifying users, making reliability a primary concern~\cite{dixit2021silent,hochschild2021cores,zeng2024soft}. 
Error propagation studies, routinely simulated through controllable fault injection experiments, reveal how injected faults corrupt intermediate values, spread during execution, and eventually affect final program outputs.
This line of research~\cite{mukherjee2003systematic,sridharan2009eliminating,fang2016epvf,li2018modeling,yang2021sugar,li2016understanding,li2017understanding,li2018modeling}, long-standing but still essential, provides insights into error behavior and continues to guide the design of effective and efficient error detection methods.
For example, Li et al.~\cite{li2018modeling} estimate instruction-level vulnerability by characterizing how errors propagate once activated, making selective instruction duplication affordable for large-scale applications.

As the key of error propagation studies, fault injection has been extensively explored in HPC~\cite{anwer2020gpu,yang2021sugar,tiwari2015understanding,huang2024versatile,rahman2021peppa,he2023demystifying,huang2023characterizing}.
Wei et al.~\cite{wei2014quantifying} modeled soft errors at both the assembly level and the LLVM intermediate representation, enabling quantifiable and accurate analysis. 
Yang et al.~\cite{yang2021enabling} leveraged GPGPU-Sim with PTXPlus mode to inject faults and capture error propagation in NVIDIA GPUs. 
Tsai et al.~\cite{tsai2021nvbitfi} introduced NVBitFI, a dynamic profiling framework built on NVBit, to support fault injection during GPU kernel execution. 
Agarwal et al.~\cite{agarwal2023resilience} proposed LLTFI, a fault injector at the LLVM IR-level, to study the trace of fault propagation in LLM.
More recently, Sun et al.~\cite{sun2025demystifying} examined LLM inference resilience from an end-to-end perspective. 
However, existing efforts remain limited: \textit{most target general-purpose HPC applications, and the few that consider LLMs evaluate error resilience only at a coarse end-to-end perspective, without providing in-depth analysis for inference.}

\begin{figure}[ht]
\centering
\includegraphics[width=1.0\columnwidth]{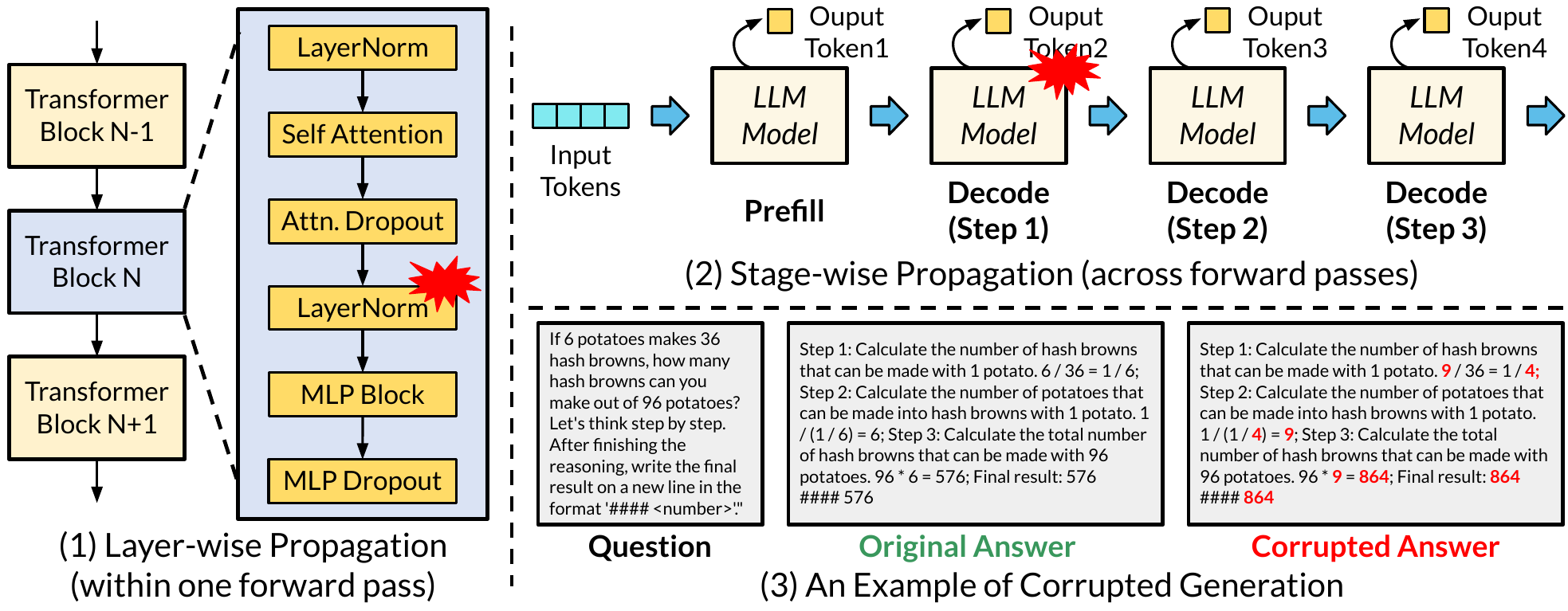}
\vspace{-6mm} 
\caption{Illustrating error propagation in LLM inference across layers (left), inference stages (middle), and a representative corrupted generation example (right).}
\label{fig:llm-error-propagation}
\end{figure}

Unlike traditional HPC workloads~\cite{alabandi2021discovering,song2022rethinking,huang2023cuszp,raje2025fastcc}, LLM inference is autoregressive: outputs are generated step by step, and each step depends on the tokens produced earlier (see Figure~\ref{fig:llm-inference}).
This nature makes error propagation in LLMs more complex, as illustrated in Figure~\ref{fig:llm-error-propagation}.
Errors can propagate at multiple levels.
At the \textit{layer-wise level} (Figure~\ref{fig:llm-error-propagation}-(1)), a fault in one submodule (e.g., layer normalization) may spread across subsequent submodules, traverse multiple Transformer blocks, and ultimately corrupt a generated token.
At the \textit{stage-wise level} (Figure~\ref{fig:llm-error-propagation}-(2)), once a generated token itself is corrupted, the autoregressive decoding process feeds this erroneous token back into LLM, causing all downstream outputs to drift and producing a corrupted final answer observable to users (Figure~\ref{fig:llm-error-propagation}-(3)).
To fully understand error propagation in LLM inference, errors must be studied at different granularities: across submodules and Transformer blocks within a forward pass, across distinct inference stages (e.g., prefill or decode iterations), and with consideration of the KV cache, which preserves conversation context and strongly influences subsequent generations.
Additional complexity arises from diverse fault types (computation vs. memory faults), diverse tasks (reasoning, math, and code generation), and diverse LLM architectures released by major AI organizations~\cite{Abdin2024Phi3TR,liu2024deepseek,team2024gemma}.
\textit{Despite these challenges, there is currently no fault injection tool that is task-agnostic, model-agnostic, location-aware, and inference-stage-aware for LLM inference, while also supporting different fault patterns.}
This gap motivates us to propose \tech and a systematic error analysis approach, which will be discussed in later sections.

\section{\tech: \textbf{L}arge \textbf{L}anguage \textbf{M}odel \textbf{F}ault \textbf{I}njector}

In this work, we propose \tech\footnote{\tech source code: \url{https://github.com/hyfshishen/LLMFI}.}, a runtime fault injector designed to understand error resilience in LLM inference.

\subsection{Overview}

\tech is designed on top of PyTorch~\cite{paszke2019pytorch} and HuggingFace Transformers~\cite{wolf2019huggingface}, ensuring broad compatibility with LLMs of different scales and architectures, from small models to large mixture-of-experts systems. \tech also supports a wide variety of inference tasks, from multi-choice reasoning~\cite{clark2018think,clark2019boolq} to code generation~\cite{chen2021evaluating}.
Additionally, for fault injection, \tech provides fine-grained control, allowing faults to be introduced at any location within the LLM model and at any timing during the inference process.

\begin{figure}[ht]
\centering
\includegraphics[width=1.0\columnwidth]{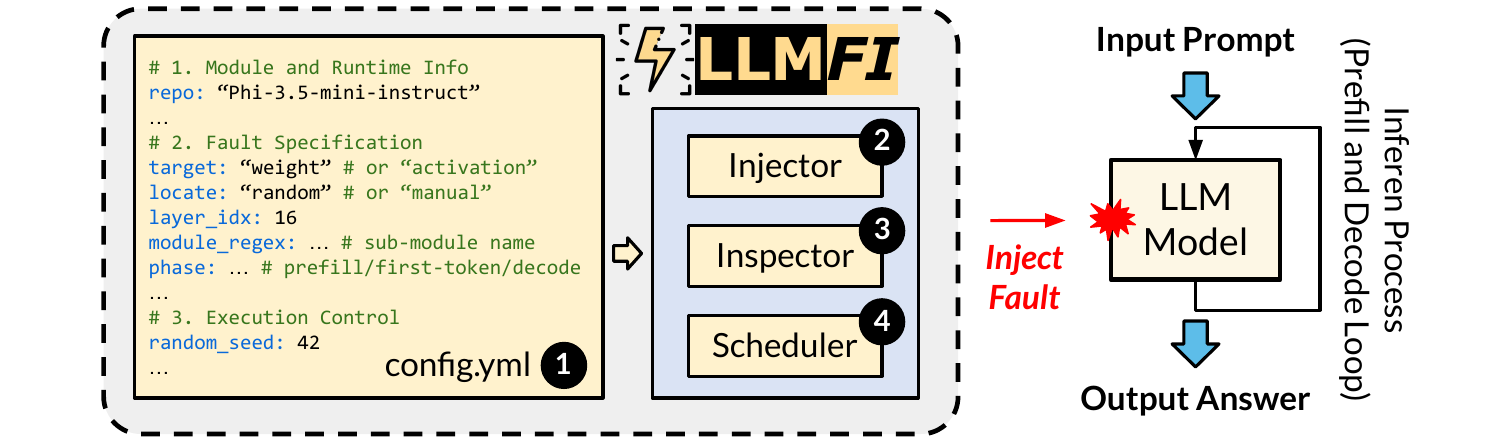}
\vspace{-4mm} 
\caption{Workflow of launching fault injection with \tech. LLMFI can inject multiple types of faults at \textit{any inference stage}, \textit{any location}, for \textit{any LLM model} and \textit{any task}.}
\label{fig:llmfi-workflow}
\end{figure}

Figure~\ref{fig:llmfi-workflow} illustrates the workflow for injecting a fault in LLM inference using \tech.
Given an arbitrary LLM model and an inference task, users first provide a configuration file (i.e., \textit{config.yml} (\ding{202})) that specifies the detailed requirements for the current fault injection trail.
Based on this configuration, \tech employs three core components to manipulate the injection process:
\textit{Injector} (\ding{203}), which defines the fault type;
\textit{Inspector} (\ding{204}), which analyzes the structure of the selected LLM model mode to determine the injection location;
\textit{Scheduler} (\ding{205}), which determines the fault injection timing.
Finally, \textit{Runtime Wrapper} integrates these components and performs the actual fault injection during the LLM inference process.


\subsection{Core Components}
\label{sec:core-componenets}

\textit{Injector} (\ding{203}) defines how bit-flip faults are performed during the runtime of LLM inference.
In this work, we assume that soft errors can occur in either computation units or memory components (see Section~\ref{sec:llm-resilience-analysis} for fault model details).
To emulate computation faults, we use {\small\texttt{forward\_hook()}} functions in PyTorch that perturb activation values immediately after a computation, mimicking faults in datapath units.
For memory faults, we employ {\small\texttt{forward\_pre\_hook()}} to corrupt weight values before they are used in computation, modeling faults in memory elements.
In both cases, bit-flips are injected in a user-controllable manner: a bit position within a numerical value can either be selected randomly or specified explicitly.
This methodology is consistent with prior fault injection approaches in machine learning dependability studies~\cite{mahmoud2020pytorchfi,sun2025demystifying,liang2025attnchecker}.

\textit{Inspector} (\ding{204}) determines the location of fault occurrence within the LLM model.
Given an arbitrary LLM model, \textit{Inspector} parses its architecture and selects a target activation or weight based on three levels of granularity: a submodule (e.g., the {\small\texttt{qkv\_proj}} projection), a functional layer (e.g., the self-attention layer), or an entire Transformer block (e.g., one of the 32 decoder layers in \textit{Phi-3.5-mini-instruct}~\cite{Abdin2024Phi3TR}).
In practice, the boundary between submodules and functional layers may vary across implementations.
For example, while complex layers like self-attention and multi-layer perceptron routinely contain multiple submodules, some LLMs~\cite{liu2024deepseek} implement layer normalization as a single Root Mean Square operation.
To accommodate such differences, \tech uses regular-expression-based matching to map user-specified target locations.
For simplicity, users can also opt to inject faults at random locations.

\textit{Scheduler} (\ding{205}) controls the timing of fault injection and supports injections at different stages of inference, including \textit{prefill}, \textit{first-token}, and any iteration of the \textit{decode loop}.
Unlike the conventional two-stage division of LLM inference into \textit{Prefill-Decode} (see Figure~\ref{fig:llm-inference}), \tech further decomposes the prefill stage into two parts: \textit{prefill}, where a forward pass over all input tokens is performed to construct the KV cache, and \textit{first-token}, where only the last input token is processed to generate the first output token.
This distinction is important because errors during \textit{prefill} primarily affect the integrity of the conversation context stored by KV cache, while errors during \textit{first-token} inference directly influence the initial output token, which is often critical for downstream generation quality and task performance~\cite{clark2018think,clark2019boolq,cobbe2021training,chen2021evaluating}.
Similarly, during the \textit{decode loop}, users can choose to inject faults at either a randomly selected iteration or a specific iteration of the decoding loop.

\subsection{Runtime Wrapper}

\begin{figure}[ht]
\centering
\includegraphics[width=1.0\columnwidth]{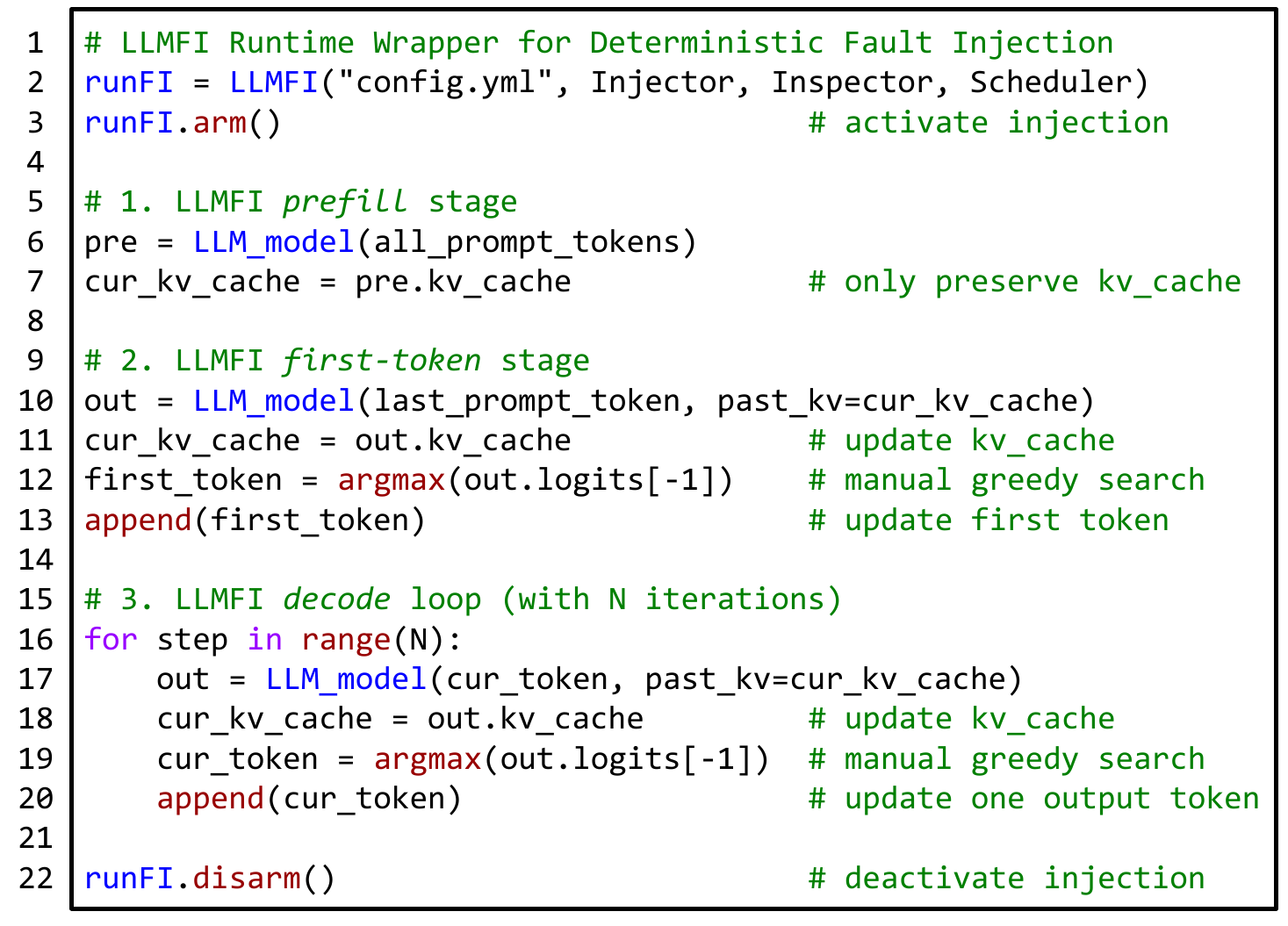}
\vspace{-6mm} 
\caption{Pseudocode of \tech \textit{Runtime Wrapper} to launch fault injection for a single LLM inference.}
\label{fig:llmfi-runtime-wrapper}
\end{figure}

Besides the configuration file and core components, \tech relies on a \textit{Runtime Wrapper} to inject faults at the specified location and timing.
The pseudocode of this design is explained in Figure~\ref{fig:llmfi-runtime-wrapper}.
After loading \textit{config.yml} and initializing the core modules (line 2), fault injection is activated by \tech.{\small\texttt{arm()}} (line 3) and will later be deactivated by \tech.{\small\texttt{disarm()}} (line 22).
While {\small\texttt{Scheduler}} determines the fault injection timing, we also manually decompose the inference process and generate output in a token-by-token manner, maintaining the KV cache and model inputs/outputs at \textit{prefill} (line 5-7), \textit{first-token} (line 9-13), and \textit{decode} loop (line 15-20).

In error propagation study, faults must be injected at different locations and timings across multiple inference trials, and the resulting output should also be compared to assess how errors propagate and ultimately affect model predictions.
This requires the underlying LLM inference process to be deterministic: given the same prompt, the model should always generate the same output.
To ensure this, we disable stochastic decoding strategies such as top-p (nucleus) sampling~\cite{ravfogel2023conformal} and temperature scaling~\cite{peeperkorn2024temperature}, and instead manually enforce greedy decoding (lines 12 and 19).
By doing so, the generated logits (i.e., raw, unnormalized output values) from each forward pass are deterministically decoded into a token with the maximum score, guaranteeing reproducible outputs and fair comparisons at the system setting and evaluation platform~\cite{yuanunderstanding}.

\subsection{Runtime Overhead of \tech}
\tech introduces an acceptable runtime overhead compared to fault-free inference, making it practical for large-scale studies. In our measurements (under the same model/task/prompt settings), the end-to-end per-prompt runtime increases by about 10\% for short generations and remains within $\sim$10-15\% for long generations. For example, on \textit{Phi-3.5-mini-instruct}~\cite{Abdin2024Phi3TR} with ARC~\cite{clark2018think}, the fault-free run takes $\sim$780-840 ms per prompt, while \tech takes $\sim$870-920 ms per prompt under the same configuration.
Across fault types, overhead differences are minor because only one fault is injected for each trial, and the injection logic is lightweight. Across stages, the overhead trend is: decode (later step) $>$ decode (early step) $\ge$ first-token $\approx$ prefill, mainly due to two reasons: (i) an extra forward pass to isolate the first-token stage for stage-aware analysis, and (ii) hook invocation for phase checking to ensure fine-grained stage-wise fault injection until the target phase is reached. Even for late-step decode injections in long generations, the overhead remains bounded (<15\% compared to fault-free executions).

\section{LLM Resilience Analysis with \tech}
\label{sec:llm-resilience-analysis}

In this section, we describe our evaluation setup and present an overall resilience analysis using \tech. We then outline several research questions on understanding error propagation, which will be further investigated and discussed in Section~\ref{sec:error-propagation}.

\subsection{Evaluation Setups}

\subsubsection{Model Selection}
We select 3 open-weighted LLMs in this work, with sizes ranging from 3.80B to 8.54B parameters (see Table~\ref{tab:llm-table}). 
These medium-scale models are developed by leading AI organizations, including Google, Microsoft, and DeepSeek AI, and have been widely evaluated in systems research~\cite{sun2025demystifying,sun2025ft2,feng2025productively,kamath2025pod,prabhu2025vattention}. 
All selected models are dense architectures; in addition, for Phi, we also include larger variants with mixture-of-experts (MoE) architectures (details in Section~\ref{sec:error-propagation}).
Beyond these core models, we also conduct supplementary evaluations on other widely used open-weight LLMs, such as \textit{Qwen2.5-7B} (7.62B)~\cite{team2024qwen2} from Alibaba, \textit{Mistral-7B-v0.3} (7.25B)~\cite{mistral-7B} from Mistral AI, and \textit{Meta-Llama-3-8B} (8.03B)~\cite{dubey2024llama} from Meta. 
The observations are consistent and highly similar to our main results, demonstrating that our findings are not specific to the 3 reported LLMs but reflect broader trends in LLM resilience.

\begin{table}[h]
\renewcommand{\arraystretch}{0.9}
\centering
\footnotesize
\caption{Details for evaluated LLM models, where \textit{Repo ID} denotes LLM's unique identifier in HuggingFace.}
\begin{tabular}{l|l|c|c|l} \toprule
{\bf Model Name} & {\bf Repo ID} & {\bf \#Params} & {\bf Release} & {\bf Publisher}\\ \midrule
Phi~\cite{Abdin2024Phi3TR}    & 
\textit{Phi-3.5-mini-instruct}  & 3.80B  &  2024-06  & Microsoft  \\
DeepSeek~\cite{bi2024deepseek} & \textit{deepseek-llm-7b-chat} &  6.91B & 2023-11 & DeepSeek AI\\
Gemma~\cite{team2024gemma} & \textit{gemma-7b} & 8.54B & 2024-02 & Google \\
\bottomrule
\end{tabular}
\label{tab:llm-table}
\end{table}

\vspace{-3mm}
\subsubsection{Task Selection}
LLMs demonstrate strong performance across a wide range of tasks. To comprehensively evaluate error impact on resilience and propagation characteristics, we conduct fault injection experiments on 13 representative tasks, summarized in Table~\ref{tab:task-table}. These tasks can be broadly grouped into four categories: reasoning~\cite{clark2018think,zellers2019hellaswag,mihaylov2018can,bisk2020piqa,sakaguchi2021winogrande,lin2021truthfulqa}, multilingual understanding~\cite{hendrycks2020measuring,ponti2020xcopa}, mathematical problem solving~\cite{cobbe2021training,chen2021evaluating}, and code generation~\cite{chen2021evaluating,austin2021program}. For inference settings such as the number of shots and the use of chain-of-thought (CoT) prompting, we strictly adopt the configuration specified in Microsoft’s Phi-3 technical report~\cite{Abdin2024Phi3TR}, which reflects industrial-level standard practices for evaluating LLM capabilities. Among these tasks, reasoning and multilingual understanding are multi-/binary-choice classification, whereas math solving and code generation are generative tasks.

\begin{table}[h]
\renewcommand{\arraystretch}{0.9}
\centering
\footnotesize
\caption{Details for evaluated tasks, where \textit{Repo ID} denotes task's unique dataset identifier in HuggingFace.}
\begin{tabular}{l|l|r|l} \toprule
{\bf Task Name} & {\bf Repo ID} & {\bf Inf. Setting} & {\bf Category}\\ \midrule
ARC~\cite{clark2018think}         & \textit{allenai/ai2\_arc} & 10-shot   & \multirow{7}{*}{Reasoning}\\
BoolQ~\cite{clark2019boolq}       & \textit{google/boolq} & 2-shot        & \\
HellaSwag~\cite{zellers2019hellaswag}   & \textit{Rowan/hellaswag} & 5-shot     & \\
OpenBookQA~\cite{mihaylov2018can}  & \textit{allenai/openbookqa} & 10-shot & \\
PIQA~\cite{bisk2020piqa}        & \textit{ybisk/piqa} & 5-shot          & \\
TruthfulQA~\cite{lin2021truthfulqa}  & \textit{truthfulqa/truthful\_qa} &    10-shot & \\ 
WinoGrande~\cite{sakaguchi2021winogrande}  & \textit{allenai/winogrande} & 5-shot  & \\\midrule
MMMLU~\cite{hendrycks2020measuring}       & \textit{openai/MMMLU} & 5-shot    & \multirow{2}{*}{Multilingual}\\
XCOPA~\cite{ponti2020xcopa}       & \textit{cambridgeltl/xcopa} & 0-shot  & \\\midrule
GSM8K~\cite{cobbe2021training}       & \textit{openai/gsm8k} & CoT, 8-shot& \multirow{2}{*}{Math}\\
MATH500~\cite{lightman2023let}     & \textit{HuggingFaceH4/MATH-500} & CoT, 0-shot& \\\midrule
HumanEval~\cite{chen2021evaluating}   & \textit{openai/openai\_humaneval} & 0-shot & \multirow{2}{*}{Code}\\
MBPP~\cite{austin2021program}        & \textit{google-research/mbpp} & 3-shot & \\
\bottomrule
\end{tabular}
\label{tab:task-table}
\end{table}

\subsubsection{Evaluation Platform}
We perform our evaluation on NVIDIA A100 GPUs (108 SMs, 40 GB) provided by the Swing cluster from Argonne National Laboratory, where each node is equipped with two AMD EPYC 7742 CPUs and 1 TB of memory, running Ubuntu 20.04 with CUDA Toolkit 12.3. The software consists of PyTorch 2.6.0 and HuggingFace Transformers 4.40.1. Inference trials are performed under several orthogonal settings, including model choice, task type, fault timing, fault location, etc, resulting in a total of $\sim$3,000 GPU hours.
Some additional experiments are also conducted on the Polaris supercomputer from Argonne National Laboratory.

\subsubsection{Fault Model}
\label{sec:fault-model}
In LLM inference, most computations are executed on the GPU side, whereas (de)tokenization on the CPU side is negligible compared to the expensive matrix multiplication.
Therefore, we inject faults on the GPU side.
Soft errors, also known as transient hardware faults, originate from hardware compute units (e.g., ALUs and flip-flops) and memory components.
We model computation faults as bit flips in intermediate activations and memory faults as bit flips in stored weights.
This fault model is consistent with state-of-the-art fault models in the HPC reliability community~\cite{sun2025demystifying,sun2025ft2,liang2025attnchecker,mahmoud2018optimizing,huang2022mitigating,huang2024versatile}.
In each trial, we inject a single bit flip.
Although single-bit errors in off-chip memory may be corrected by ECC, this model remains meaningful because (i) not all GPUs or on-chip resources provide full ECC protection~\cite{park2024cachecraft}, and (ii) the single-bit model represents the minimal perturbation needed to study error resilience and propagation. 
Multi-bit cases for memory faults are further explored in Section~\ref{sec:error-propagation}.

\begin{table*}[h]
\renewcommand{\arraystretch}{0.9}
\centering
\footnotesize
\caption{Error resilience results for 3 LLMs on 13 tasks using \tech. For each model and task, we inject computation and memory faults at three inference stages: \textit{Prefill}, \textit{First-Token}, and \textit{Decode}. We report the resulting accuracy, where \textit{Ori.}, \textit{Com.}, and \textit{Mem.} denote the accuracy of fault-free runs, runs with computation faults, and runs with memory faults, respectively. For example, in the ARC task with Phi, accuracy is 81.76 when computation faults are injected during the first-token stage.}
\label{tab:resilience-evaluation-table}
\begin{tabular}{l|c|l|lll|lll|lll|lll} \toprule
\multirow{2}{*}{\bf Task Name} & \multirow{2}{*}{\bf Task Details} & \multirow{2}{*}{\bf Inference Stage} & \multicolumn{3}{c|}{\bf Phi} & \multicolumn{3}{c|}{\bf DeepSeek} & \multicolumn{3}{c|}{\bf Gemma} & \multicolumn{3}{c}{\bf Average} \\\cmidrule(lr){4-6}\cmidrule(lr){7-9}\cmidrule(lr){10-12}\cmidrule(lr){13-15}
& & & \textit{Ori.} & \textit{Com.}  & \textit{Mem.} & \textit{Ori.} & \textit{Com.}  & \textit{Mem.} & \textit{Ori.} & \textit{Com.}  & \textit{Mem.} & \textit{Ori.} & \textit{Com.}  & \textit{Mem.}\\\midrule
\multirow{3}{*}{\bf ARC} & 10-shot& \textit{Prefill} &  \multirow{3}{*}{85.20} & 84.08 & 83.82 & \multirow{3}{*}{62.31} & 62.13 & 61.02 & \multirow{3}{*}{81.41} & 80.98 & 79.95 & \multirow{3}{*}{76.31} & 75.73 & 74.93 \\
& Reasoning & \textit{First-Token} & & 81.76 & 84.34 &  & 59.98 & 62.13 &  & 78.49 & 80.98 & & 73.41 & 75.82 \\
& Multi-choice & \textit{Decode} & & 85.03 & 85.20 &  & 62.22 & 62.31 &  & 81.41 & 81.41 & & 76.22 & 76.31 \\\midrule
\multirow{3}{*}{\bf BoolQ} & 2-shot & \textit{Prefill} & \multirow{3}{*}{84.00} & 82.71 & 81.67 & \multirow{3}{*}{82.13} & 81.40 & 80.66 & \multirow{3}{*}{85.01} & 84.33 & 83.81 & \multirow{3}{*}{83.71} & 82.81 & 82.05 \\
& Reasoning & \textit{First-Token} & & 79.35 & 83.11 & & 77.75 & 81.79 & & 81.30 & 84.46 & & 79.47 & 83.12 \\
& Multi-choice & \textit{Decode}  & & 84.00 & 84.00 & & 82.13 & 82.13 & & 85.01 & 85.01 & & 83.71 & 83.71 \\\midrule
\multirow{3}{*}{\bf HellaSwag} & 5-shot & \textit{Prefill} & \multirow{3}{*}{73.68} & 72.62 & 71.92 & \multirow{3}{*}{59.78} & 59.19 & 58.46 & \multirow{3}{*}{67.60} & 67.02 & 66.31 & \multirow{3}{*}{67.02} & 66.28 & 65.56 \\
& Reasoning &\textit{First-Token} & & 71.31 & 73.21 & & 57.37 & 59.36 & & 64.79 & 67.26 & & 64.49 & 66.61 \\
& Multi-choice & \textit{Decode} & & 73.65 & 73.70 & & 59.70 & 59.76 & & 67.53 & 67.60 & & 66.96 & 67.02 \\\midrule
\multirow{3}{*}{\bf OpenBookQA} & 10-shot & \textit{Prefill} & \multirow{3}{*}{80.20} & 78.98 & 77.35 & \multirow{3}{*}{66.53} & 65.71 & 65.71 & \multirow{3}{*}{81.84} & 81.22 & 79.80 & \multirow{3}{*}{76.19} & 75.30 & 74.29 \\
& Reasoning & \textit{First-Token} & & 77.76 & 79.18 & & 64.49 & 65.71 & & 79.18 & 81.63 & & 73.81 & 75.51 \\
& Multi-choice & \textit{Decode} & & 80.20 & 80.20 & & 66.53 & 66.53 & & 81.84 & 81.84 & & 76.19 & 76.19 \\\midrule
\multirow{3}{*}{\bf PIQA} & 5-shot & \textit{Prefill} & \multirow{3}{*}{82.82} & 82.05 & 81.45 & \multirow{3}{*}{67.16} & 66.07 & 66.23 & \multirow{3}{*}{82.76} & 81.72 & 81.67 & \multirow{3}{*}{77.58} & 76.61 & 76.45 \\
& Reasoning & \textit{First-Token} & & 78.89 & 82.27 & & 64.54 & 66.83 & & 79.54 & 82.49 & & 74.32 & 77.20 \\
& Multi-choice & \textit{Decode} & & 82.71 & 82.87 & & 67.05 & 67.16 & & 82.65 & 82.71  & & 77.47 & 77.58 \\\midrule
\multirow{3}{*}{\bf TruthfulQA} & 10-shot & \textit{Prefill} & \multirow{3}{*}{52.79} & 52.29 & 52.29 & \multirow{3}{*}{51.67} & 51.30 & 50.93 & \multirow{3}{*}{83.89} & 83.77 & 83.27 & \multirow{3}{*}{62.78} & 62.45 & 62.16 \\
& Reasoning & \textit{First-Token} & & 51.05 & 52.29 & & 48.95 & 51.55 & & 81.04 & 83.74 & & 60.35 & 62.53 \\
& Multi-choice & \textit{Decode} & & 52.66 & 52.79 & & 51.67 & 51.67 & & 83.77 & 83.89 & & 62.70 & 62.78 \\\midrule
\multirow{3}{*}{\bf WinoGrande} & 5-shot & \textit{Prefill} & \multirow{3}{*}{70.05} & 69.49 & 69.02 & \multirow{3}{*}{53.72} & 52.61 & 53.33 & \multirow{3}{*}{64.74} & 63.47 & 64.50 & \multirow{3}{*}{62.84} & 61.86 & 62.28 \\
& Reasoning & \textit{First-Token} & & 66.80 & 69.65 & & 51.98 & 53.57 & & 62.28 & 64.74 & & 60.35 & 62.65 \\
& Multi-choice & \textit{Decode} & & 70.05 & 70.05 & & 53.49 & 53.72 & & 64.66 & 64.74 & & 62.73 & 62.84 \\\midrule
\multirow{3}{*}{\bf MMMLU} & 5-shot & \textit{Prefill} & \multirow{3}{*}{46.08} & 45.97 & 45.94 & \multirow{3}{*}{34.91} & 34.75 & 33.92 & \multirow{3}{*}{49.91} & 49.85 & 49.36 & \multirow{3}{*}{43.63} & 43.52 & 43.07 \\
& Multilingual & \textit{First-Token} & & 45.73 & 46.01 & & 34.46 & 33.83 & & 49.22 & 49.73 & & 43.14 & 43.19 \\
& Multi-choice & \textit{Decode} & & 46.08 & 46.08 & & 34.91 & 34.87 & & 49.91 & 49.91 & & 43.63 & 43.62 \\\midrule
\multirow{3}{*}{\bf XCOPA} & 0-shot & \textit{Prefill} & \multirow{3}{*}{59.67} & 59.57 & 59.62 & \multirow{3}{*}{54.26} & 51.91 & 52.74 & \multirow{3}{*}{53.22} & 52.22 & 53.17 & \multirow{3}{*}{55.72} & 54.57 & 55.18 \\
& Multilingual & \textit{First-Token} & & 58.79 & 59.55 & & 52.08 & 54.16 & & 52.24 & 53.19 & & 54.37 & 55.63 \\
& Multi-choice & \textit{Decode} & & 59.67 & 59.59 & & 54.25 & 54.18 & & 53.22 & 53.22 & & 55.71 & 55.66 \\\midrule
\multirow{3}{*}{\bf GSM8K} & CoT, 8-shot & \textit{Prefill} & \multirow{3}{*}{83.03} & 82.92 & 80.72 & \multirow{3}{*}{52.87} & 52.10 & 52.22 & \multirow{3}{*}{47.76} & 46.15 & 44.93 & \multirow{3}{*}{61.22} & 60.39 & 59.29 \\
& Math & \textit{First-Token} & & 82.90 & 80.22 & & 52.47 & 51.60 & & 44.55 & 45.23 & & 59.97 & 59.02 \\
& Generative & \textit{Decode} & & 82.91 & 82.54 & & 52.87 & 52.63 & & 45.63 & 45.51 & & 60.47 & 60.23 \\\midrule
\multirow{3}{*}{\bf MATH500} & CoT, 0-shot & \textit{Prefill} & \multirow{3}{*}{35.94} & 34.83 & 34.49 & \multirow{3}{*}{13.07} & 12.50 & 11.86 & \multirow{3}{*}{20.97} & 19.46 & 18.41 & \multirow{3}{*}{23.33} & 22.26 & 21.59 \\
& Math & \textit{First-Token} & & 35.04 & 34.01 & & 11.22 & 11.79 & & 18.99 & 18.20 & & 21.75 & 21.33 \\
& Generative & \textit{Decode} & & 35.14 & 34.94 & & 11.90 & 11.67 & & 19.20 & 18.47 & & 22.08 & 21.69 \\\midrule
\multirow{3}{*}{\bf HumanEval} & 0-shot & \textit{Prefill} & \multirow{3}{*}{54.27} & 52.44 & 50.61 & \multirow{3}{*}{40.81} & 38.28 & 36.59 & \multirow{3}{*}{32.22} & 29.81 & 27.48 & \multirow{3}{*}{42.43} & 40.18 & 38.23 \\
& Code & \textit{First-Token} & & 50.00 & 49.39 & & 37.21 & 35.98 & & 28.24 & 26.91 & & 38.48 & 37.43 \\
& Generative & \textit{Decode} & & 53.82 & 53.42 & & 40.03 & 40.12 & & 32.12 & 32.01 & & 41.99 & 41.85 \\\midrule
\multirow{3}{*}{\bf MBPP} & 3-shot & \textit{Prefill} & \multirow{3}{*}{44.24} & 43.03 & 42.78 & \multirow{3}{*}{31.52} & 30.43 & 30.37 & \multirow{3}{*}{33.49} & 32.12 & 30.30 & \multirow{3}{*}{36.42} & 35.19 & 34.48 \\
& Code & \textit{First-Token} & & 43.64 & 42.42 & & 29.21 & 29.97 & & 30.91 & 29.09 & & 34.59 & 33.83 \\
& Generative & \textit{Decode} & & 44.16 & 44.11 & & 30.92 & 30.76 & & 33.42 & 33.13 & & 36.17 & 36.00 \\
\bottomrule
\end{tabular}
\end{table*}

\subsection{Evaluation Approach}
\label{sec:resilience-analysis-approach}

We perform fault injection using \tech and evaluate error resilience for LLM inference across different models and tasks. 
Then, we summarize key \textbf{takeaways} and propose related research questions that guide our analysis of error propagation in Section~\ref{sec:error-propagation}. 
For multi-/binary-choice reasoning tasks~\cite{clark2018think,zellers2019hellaswag,mihaylov2018can,bisk2020piqa,sakaguchi2021winogrande,lin2021truthfulqa,hendrycks2020measuring,ponti2020xcopa}, we measure accuracy as the ratio of correct predictions to the total number of predictions.
For math tasks~\cite{cobbe2021training,chen2021evaluating}, answers are extracted using regular expression–based parsing and the official evaluation scripts (exact-match and answer-check harnesses) provided by benchmark authors. For code generation tasks~\cite{chen2021evaluating,austin2021program}, we adopt the standard execution-based evaluation harnesses, in which generated programs are executed against benchmark unit tests to determine functional correctness. 
All generative tasks are also reported in terms of accuracy for consistency. 
By comparing the accuracy of golden runs against that of fault injections, users can quantify the impact of injected faults on LLM inference.
To ensure that the results reflect each model’s actual inference capability, we fine-tune prompts for every task and model. The golden-run inference accuracy aligns with the numbers reported in official technical reports~\cite{Abdin2024Phi3TR,bi2024deepseek,team2024gemma,team2024qwen2}. Although several tasks (e.g., MBPP on Phi) yield slightly lower accuracy due to unavailable closed-source prompt templates, this does not affect the fairness of our analysis, since both golden runs and fault-injection trials are evaluated under identical conditions.
Unless otherwise stated, inference is conducted in {\small\texttt{float16}}, with other data types discussed in Section~\ref{sec:error-propagation}.

\subsection{Resilience Analysis}
\label{sec:resilience-analysis}

Table~\ref{tab:resilience-evaluation-table} presents the fine-grained error resilience of 3 LLMs across 13 tasks. 
For each task and LLM model, under a given fault setting (computation or memory fault and inference stage), we randomly select a location within a submodule of an arbitrary Transformer block for injection, and repeat 10 times per input data sample. 
For decode-stage injections, we additionally sample a random loop iteration, while the maximum generation length for each task strictly follows the industrial setting~\cite{Abdin2024Phi3TR}.
As seen, \textbf{Takeaway 1: Fault injection reveals measurable accuracy degradation across all tasks, with accuracy reductions of up to 4.25\% in multi-choice reasoning and 5.01\% in generative tasks.}
Even small reductions are notable in safety-critical applications, underscoring the need to study error propagation in LLM inference.

Another important observation across all tasks is that \textbf{Takeaway 2: The inference stage at which a fault occurs has a dominant impact on error resilience.} For reasoning tasks, which are typically formulated as multi-choice questions~\cite{clark2019boolq,mihaylov2018can,sakaguchi2021winogrande}, the answer is constrained to a fixed output format (e.g., ending with \textit{“Answer: X”}). In this setting, only the first generated token determines correctness, meaning that faults in the prefill or first-token stage affect accuracy, whereas faults in later decode steps do not. 
In practice, decode-stage faults do cause negligible fluctuations; for example, in PIQA on Phi, accuracy changes slightly from 82.82 to 82.71 (memory) or 82.78 (computation) (<0.1\% difference), which is attributable to noise in the answer-extraction process rather than meaningful semantic errors.
We further find that \textbf{Takeaway 3: First-token stage in LLM inference is highly sensitive to computation faults but less affected by memory faults}. Computation faults directly perturb activations that immediately determine the first generated token, which in reasoning tasks often corresponds to the final answer. In contrast, memory faults perturb weights, whose effects take time to manifest. During first-token generation, the model processes only the last prompt token, providing limited opportunities for memory faults to influence the output.
Since first-token generation is especially fragile to computation faults, we focus on this stage to guide our deeper analysis -- \textit{Section~\ref{sec:layer-wise-error-propagation}: How do errors propagate across different layers within an LLM?}

In Table~\ref{tab:resilience-evaluation-table}, we also observe that \textbf{Takeaway 4: Generative tasks End-to-end error resilience analysis is insufficient for generative tasks}, including math solving (GSM8K~\cite{cobbe2021training} and MATH500~\cite{lightman2023let}) and code generation (HumanEval~\cite{chen2021evaluating} and MBPP~\cite{austin2021program}). Unlike reasoning tasks, generative tasks typically require relatively long output sequences to construct complete answers, regardless of whether chain-of-thought (CoT) prompting is used. In Table~\ref{tab:resilience-evaluation-table}, faults are injected into a single randomly selected decode step, which is not enough to capture how errors propagate across multiple steps and influence the final output. 
Moreover, taking HumanEval benchmark as an example, our inference setting allows up to 512 output tokens, ensuring coverage for all samples, but also meaning that faults may be injected into irrelevant positions for tasks with much shorter answers.
This motivates a deeper sample-level study, leading to \textit{Section~\ref{sec:stage-wise-error-propagation}: How do errors propagate across decode steps?}

In Table~\ref{tab:resilience-evaluation-table}, we can find that \textbf{Takeaway 5: Reliable error resilience analysis requires strong baseline performance; otherwise inference outcomes are dominated by noise.} This limitation is evident in multilingual benchmarks such as MMMLU~\cite{hendrycks2020measuring} and XCOPA~\cite{ponti2020xcopa}. In MMMLU, accuracy under computation and memory faults during first-token generation is nearly identical, contradicting Takeaway 3, simply because the model performs poorly on non-English questions. Similarly, XCOPA emphasizes low-resource languages (e.g., Estonian, Haitian Creole), where selected LLMs achieve accuracy close to random guessing. In these cases, injected faults have little measurable effect, highlighting that resilience results are only meaningful when the baseline task accuracy is sufficiently reliable. This also validates our methodology of reporting accuracy task by task and adhering to industry-standard prompts.
Beyond this, we also observe an intriguing phenomenon in DeepSeek: the model sometimes switches its output language from English to Chinese -- the two dominant languages in DeepSeek's training corpus~\cite{liu2024deepseek}. This suggests that fault-induced behavior may manifest not only as accuracy loss but also as cross-language drift. 

Recall that in the \textit{Scheduler} of \tech, we intentionally decouple the original prefill into two stages: \textit{prefill} and \textit{first-token} generation. 
Faults in the first-token stage directly affect the generation of the initial output token, thereby influencing subsequent decoding, whereas faults in the prefill stage primarily corrupt the KV cache. 
Computation and memory faults corrupt either activations or model weights, both of which can distort context stored in the KV cache.
\textbf{Takeaway 6: In reasoning tasks, computation and memory faults in the prefill stage lead to comparable accuracy degradation due to distorted KV cache, but in generative tasks, memory faults have a greater impact.}
Specifically, except for the polluted initial KV cache, computation faults affect only a single forward pass in generative tasks that produce long sequences: if the error does not alter the current token, subsequent decoding proceeds normally without any further error propagation. In contrast, memory faults continuously perturb LLM weights across the entire inference phase, leading to sustained degradation (e.g., average accuracy of 40.18 vs. 38.23 in HumanEval).
Beyond weights, inference also relies on GPU-resident KV cache buffers, which can themselves be susceptible to soft errors. This raises an additional evaluation perspective: \textit{Section~\ref{sec:memory-kv-cache}: What is the impact of soft errors that directly corrupt the memory storing KV cache?}

Beyond this, several additional dimensions are essential for understanding error propagation in LLM inference.
First, all experiments use {\small\texttt{float16}} arithmetic; different numeric precisions may exhibit distinct sensitivities to bit flips, motivating us to understand the impact of numeric precision and bit-flip location \textit{(Section~\ref{sec:bit-flip-location-data-type})}.
Second, our evaluated models are dense architectures, while modern LLMs increasingly adopt mixture-of-experts (MoE) designs with sparse expert activation, raising an investigation into error propagation behaviors for MoE models \textit{(Section~\ref{sec:dense-or-moe})}. 
Finally, we extend our single-bit analysis to consider multi-bit upsets in \textit{Section~\ref{sec:multiple-bit-flips}}.
Together, these perspectives complete the analysis of this work.

\section{Error Propagation Analysis with \tech}
\label{sec:error-propagation}

In this section, we perform a deeper error propagation study and analysis using \tech from the following perspectives.

\subsection{Layer-wise Error Propagation}
\label{sec:layer-wise-error-propagation}

As discussed in Section~\ref{sec:resilience-analysis}, the first-token stage is particularly vulnerable to computation faults in reasoning tasks.
We therefore use this stage to analyze how errors propagate across different layers of the model.
Here, each layer refers to a Transformer block that includes a multi-layer perceptron (MLP), self-attention module, layer normalization, dropout, and activation functions (mostly \texttt{\small SiLU}).
Among the three selected LLMs, Phi, DeepSeek, and Gemma, their architectures contain 32, 30, and 28 Transformer blocks, respectively~\cite{Abdin2024Phi3TR,bi2024deepseek,team2024gemma}.
For fair comparison across layers, we fix the fault injection configuration (i.e., submodule, activation site, and bit index) to control variables.
The injection methodology follows that described in Section~\ref{sec:resilience-analysis-approach}.
We evaluate on the ARC and BoolQ benchmarks, as they are representative reasoning tasks that exhibit consistent trends observed in other datasets.

\begin{figure}[ht]
\centering
\includegraphics[width=1.0\columnwidth]{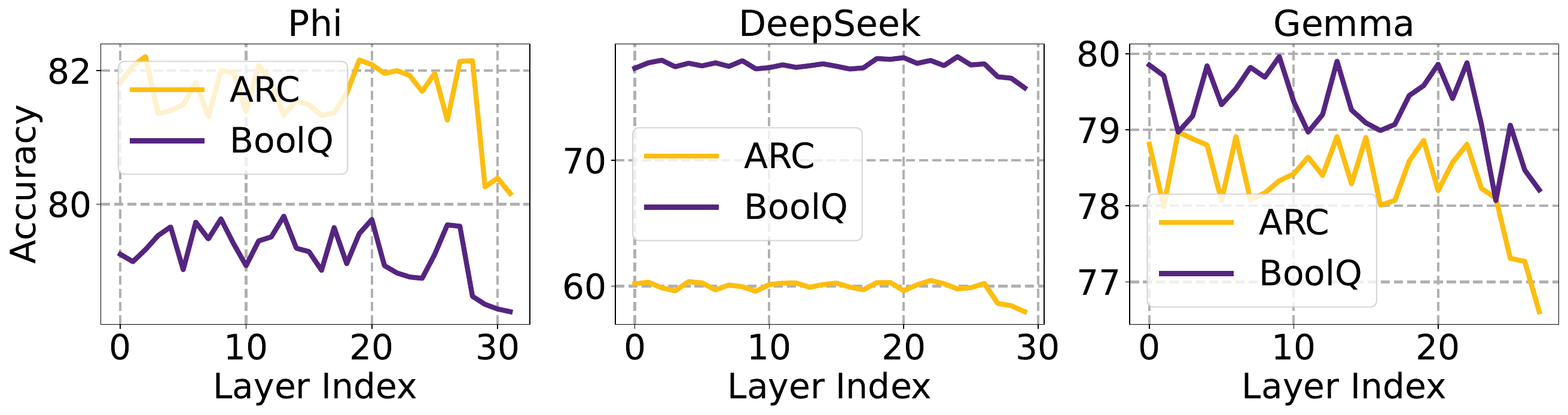}
\vspace{-4mm} 
\caption{Exploring error impacts on different Transformer block layers using ARC and BoolQ benchmarks with 3 LLMs.}
\label{fig:error-propagation-transformer-block}
\end{figure}

Figure~\ref{fig:error-propagation-transformer-block} presents the accuracy under computation faults injected into different Transformer blocks.
As shown, \textbf{Takeaway 7: Accuracy remains stable across early and middle layers but degrades consistently in the final 3 to 4 Transformer blocks.}
In ARC on Phi, the first 29 blocks maintain accuracy between 81.26 and 82.21, whereas the last three blocks drop to 80.16 to 80.39.
This trend is not limited to reasoning tasks; similar patterns are observed during first-token generation in generative workloads.
Additionally, \textbf{Takeaway 8: Transformer layers exhibit a natural self-correction effect, where intermediate computations can mitigate faults.}
In early and middle layers, nonlinear activations and residual connections collaboratively re-normalize corrupted activations, preventing errors from propagating further.
However, the final layers lack sufficient downstream correction, allowing accumulated distortions to directly influence the output logits and subsequently the generated tokens.
While prior work~\cite{sun2025ft2} highlights the corrective role of nonlinear activations, we observe that linear transformations can also partially correct errors.

\begin{figure}[ht]
\centering
\includegraphics[width=1.0\columnwidth]{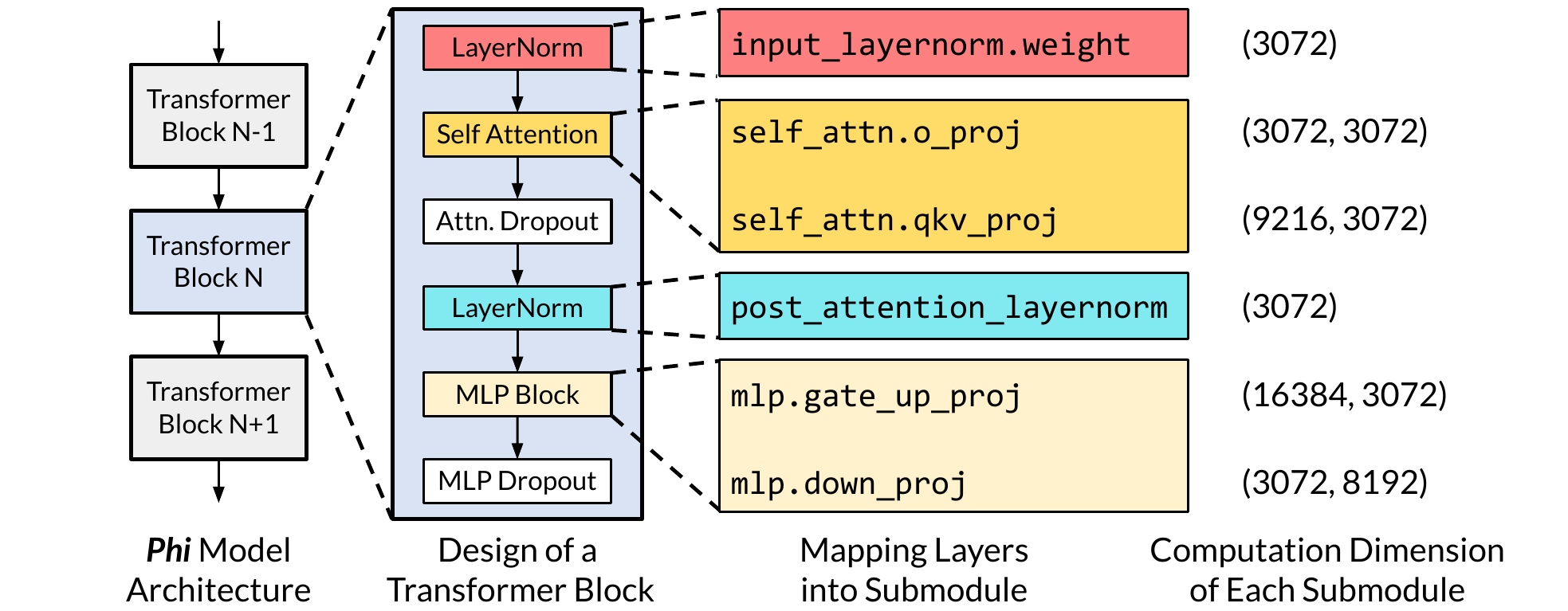}
\vspace{-4mm} 
\caption{Submodules in a Transformer block of Phi.}
\label{fig:submodule-details-of-phi}
\end{figure}

To pinpoint which linear submodules mitigate soft errors within a Transformer block, we first review the architecture of modern LLM implementations~\cite{wolf2019huggingface}.
An example from Phi is illustrated in Figure~\ref{fig:submodule-details-of-phi}, where activation functions are omitted for clarity.
Each Transformer block begins with a \texttt{\small layer\_norm} operation, which normalizes input statistics without changing tensor dimensions.
It is followed by a \texttt{\small self\_attention} module composed of \texttt{\small qkv\_proj} and \texttt{\small o\_proj} linear layers that perform query–key–value projection and output transformation, respectively.
After attention, another \texttt{\small layer\_norm} and \texttt{\small dropout} appear to stabilize training and prevent overfitting.
A subsequent MLP block contains two key projections: \texttt{\small gate\_up\_proj} and \texttt{\small down\_proj}, which expand and then contract the hidden dimension to facilitate nonlinear feature transformation.
Note that \texttt{\small dropout} is inactive during inference and therefore does not influence error propagation in our evaluation.
DeepSeek and Gemma adopt nearly identical Transformer block structures; the main differences lie in their projection layers (e.g., \texttt{\small down\_proj} of size (3072, 24576) in Gemma) and implementation details, such as separating \texttt{\small qkv\_proj} and \texttt{\small gate\_up\_proj} into distinct linear operations.

\begin{figure}[ht]
\centering
\includegraphics[width=1.0\columnwidth]{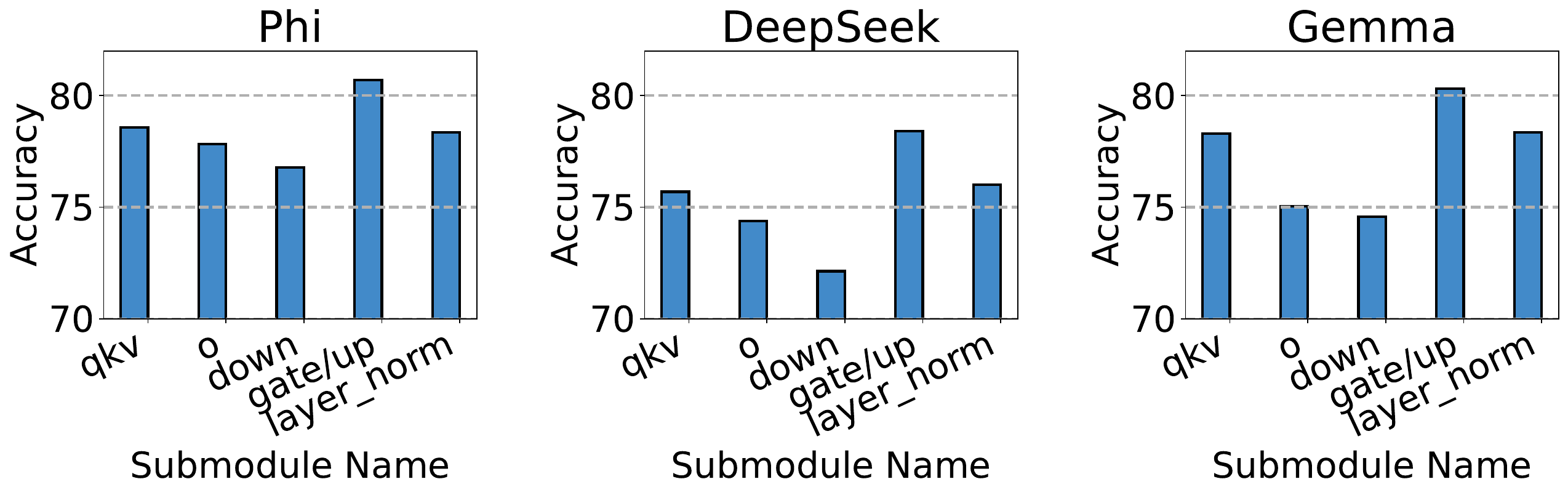}
\vspace{-6mm} 
\caption{Investigating error impacts on different submodules within the last Transformer block using BoolQ benchmarks.}
\label{fig:error-propagation-within-submodule}
\end{figure}

Figure~\ref{fig:error-propagation-within-submodule} further presents submodule-level error resilience using the BoolQ task.
Among them, \texttt{\small gate\_up\_proj} shows the highest error resilience.
In contrast, both \texttt{\small qkv\_proj} and \texttt{\small layer\_norm} demonstrate moderate resilience, and whether they reside within or outside of self-attention computations has limited influence on error resilience.
Surprisingly, \texttt{\small down\_proj} consistently emerges as the most error-prone submodule, which has accuracy only 76.78, 72.13, and 74.58 for Phi, DeepSeek, and Gemma, respectively.
To understand this pattern, we correlate resilience with the dimensionality in each submodule.
\textbf{Takeaway 9: Dimensional expansion mitigates the impact of soft errors, whereas dimensional reduction amplifies it.}
Specifically, \texttt{\small gate\_up\_proj} expands the hidden dimension (e.g., from length 3072 to 16384 in Phi), which spreads bit-flip-induced perturbations over a larger numerical space, effectively averaging out local deviations.
In contrast, \texttt{\small down\_proj} compresses high-dimensional activations back to the model’s base hidden size, concentrating accumulated numerical noise.
This dimensionality-dependent behavior also explains why most Transformer layers can self-correct errors besides non-linear activation functions.

\subsection{Stage-wise Error Propagation}
\label{sec:stage-wise-error-propagation}

As discussed in Section~\ref{sec:resilience-analysis}, end-to-end resilience analysis is insufficient for generative tasks. 
We use \tech to inject faults across different iterations within the decode loop and analyze stage-wise error propagation during inference. 
We focus on GSM8K~\cite{cobbe2021training} and HumanEval~\cite{chen2021evaluating} as representative tasks for math solving and code generation, and conduct experiments using the Phi model. 
Both computation and memory faults are injected into decode iterations at steps 2, 4, 8, 16, 32, and 64, while other fault configurations (e.g., injection site and bit index) are fixed for consistency.

\begin{figure}[ht]
\centering
\includegraphics[width=1.0\columnwidth]{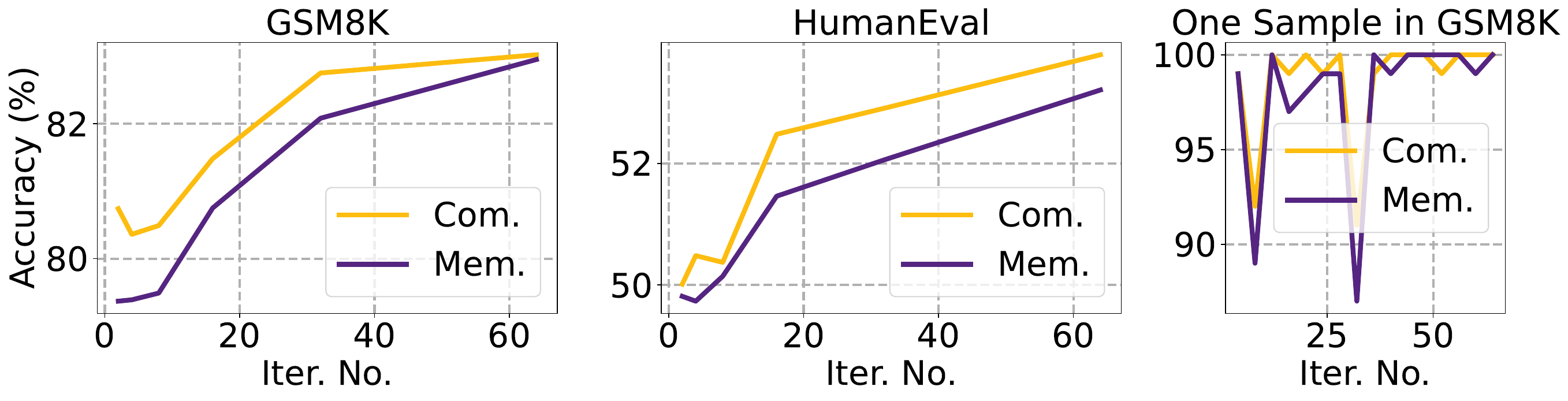}
\vspace{-6mm} 
\caption{Exploring impacts on iterations within decode loops using Phi on generative GSM8K and HumanEval tasks.}
\label{fig:error-propagation-different-decode-loop}
\end{figure}

Figure~\ref{fig:error-propagation-different-decode-loop} (left and middle) shows the results for GSM8K and HumanEval, respectively. 
\textbf{Takeaway 10: Errors injected in early decode iterations are less likely to recover than those in later steps, and memory faults are marginally more disruptive than computation faults.} 
This variability stems from the heterogeneous semantic contribution of each generated token; early tokens often define problem context or reasoning trajectory, while later tokens tend to refine already established logic. 
To further understand this, we analyze a representative GSM8K sample by injecting 100 faults every four iterations between steps~4 and~64.
Similarly, we align other fault settings. 
Results are in Figure~\ref{fig:error-propagation-different-decode-loop} (right).
Interestingly, while most iterations maintain stable accuracy (around 100\%), two specific decoding steps exhibit up to 10\% accuracy degradation. 
This phenomenon is also observed in other input samples. 
\textbf{Takeaway 11: In generative tasks, certain decode steps (or tokens) act as critical points that disproportionately influence accuracy.} 
These points likely correspond to tokens that anchor the reasoning chain or delimit logical transitions, such as numerical values involved in computation rather than function words (e.g., ``the"), making them more vulnerable to transient perturbations. 
Identifying such sensitivity patterns across datasets and models is non-trivial and represents a promising direction for future work.

\subsection{Direct Memory Faults in KV Cache}
\label{sec:memory-kv-cache}

In \tech, computation faults naturally propagate into the KV cache through corrupted activations during normal execution. 
However, prior memory fault experiments in Table~\ref{tab:resilience-evaluation-table} only targeted model weights, omitting the GPU memory region that stores the KV cache itself.
As the inference sequence grows, the KV cache expands proportionally~\cite{liu2024cachegen}, becoming a large component of GPU memory. 
To study this, we extend \tech’s \textit{Runtime Wrapper} to support direct fault injection into the KV cache region, integrating the same memory-level injection logic used in the \textit{Injector}. 
We evaluate this behavior using the ARC benchmark for clarity; similar trends are observed across other reasoning and generative tasks.

\begin{figure}[ht]
\centering
\includegraphics[width=1.0\columnwidth]{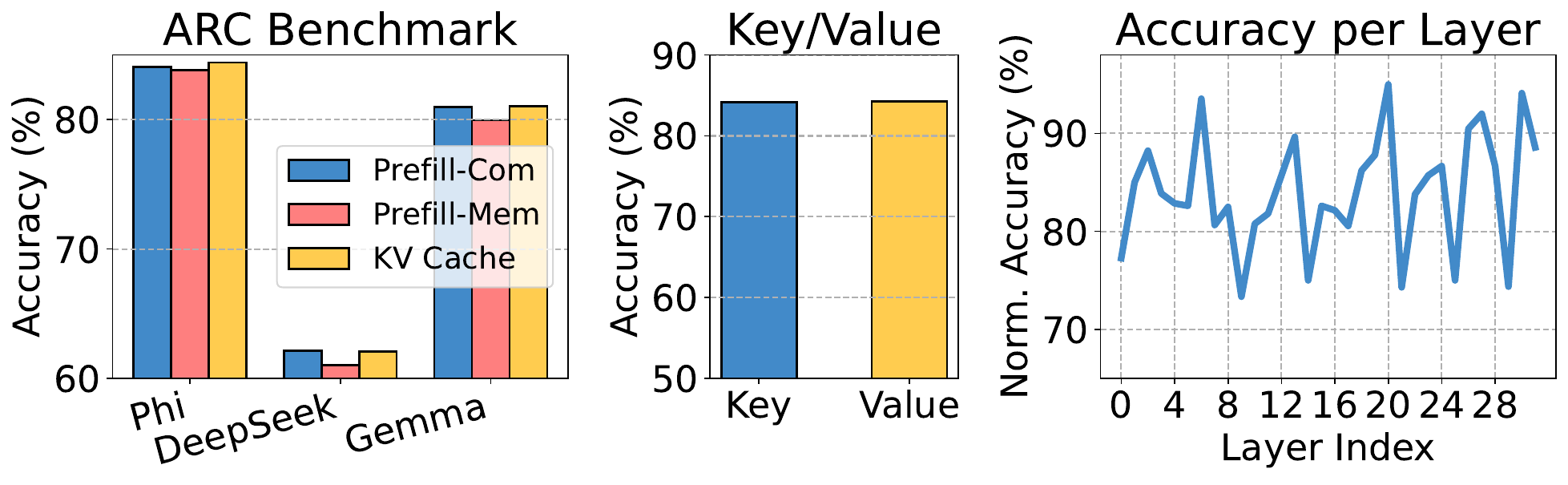}
\vspace{-7mm} 
\caption{Evaluation of KV cache memory faults using ARC, with fine-grained analysis (middle and right) using Phi.}
\label{fig:memory-fault-in-kv-cache}
\end{figure}

Figure~\ref{fig:memory-fault-in-kv-cache} (left) compares the accuracy under computation faults in prefill, indirect memory faults (propagated through activations, still in prefill), and direct memory faults on the KV cache. 
As shown, direct memory faults lead to modest accuracy reductions, ranging from 0.23 in DeepSeek to 0.78 in Phi, indicating that \textbf{Takeaway 12: The KV cache is generally more resilient to memory faults than LLM model weights.} 
This trend is consistent across both key and value arrays (84.16 vs 84.28 accuracy), as illustrated in Figure~\ref{fig:memory-fault-in-kv-cache} (middle).
When investigating KV cache at layer granularity, we uncover a distinct pattern: \textbf{Takeaway 13: KV cache sensitivity fluctuates across Transformer layers, suggesting that resilience is determined by internal attention semantics rather than layer depth alone.} 
In Figure~\ref{fig:memory-fault-in-kv-cache} (right), for instance, while layer~8 maintains 82.50 accuracy, layer~9 in the same model drops sharply to 73.33. 
This oscillation likely reflects differences in how layers contribute to context formation.
KV cache in layers that aggregate global dependencies or encode key semantics appears more sensitive to memory faults.
The pattern varies across tasks and models, and its relation to attention head activity remains an open question for future study.

\subsection{Impact of Bit-flip Location and Data Type}
\label{sec:bit-flip-location-data-type}

We examine how the bit-flip location affects error propagation. 
Similarly, we use the ARC and BoolQ benchmarks under computation faults injected during the prefill stage. 

\begin{figure}[ht]
\centering
\includegraphics[width=1.0\columnwidth]{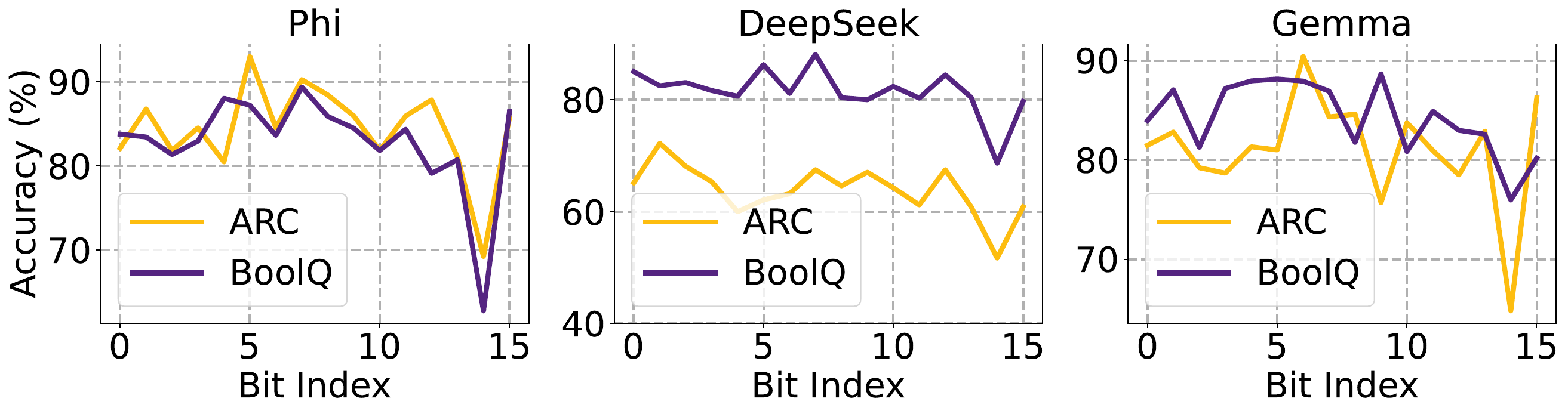}
\vspace{-6mm} 
\caption{Accuracy with the bit index soft error occurs.}
\label{fig:understanding-impact-of-bit-location}
\end{figure}

Figure~\ref{fig:understanding-impact-of-bit-location} summarizes the results. 
Taking BoolQ on Phi as an example, a flip at bit index~14 yields only 62.79 accuracy, whereas flips at other bit positions result in accuracy from 80.46 to 92.89. 
This trend is consistent across ARC, as well as DeepSeek and Gemma. 
\textbf{Takeaway 14: In \texttt{\small float16} inference, most bit flips can be attenuated, but bit 14 in exponent section cause catastrophic deviations that propagate irrecoverably.} 
Bit 14 represents the most significant exponent bit in \texttt{\small float16} format; flipping it exponentially scales the affected value, producing large-magnitude activations.

\begin{figure}[ht]
\centering
\includegraphics[width=1.0\columnwidth]{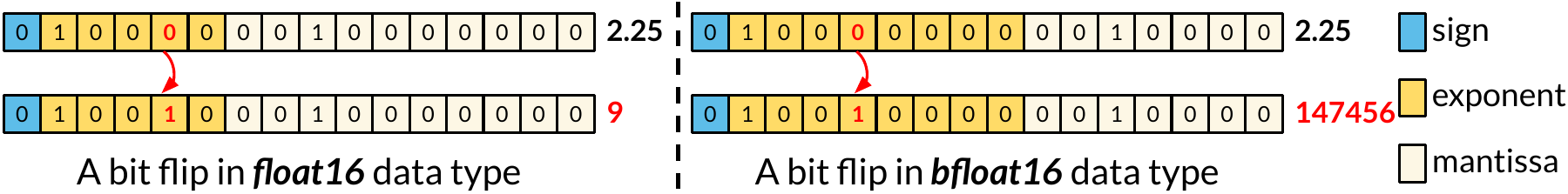}
\vspace{-6mm} 
\caption{Illustrating why \texttt{\small bfloat16} is more vulnerable to bit flips than \texttt{\small float16} in LLM inference.}
\label{fig:illustrating-why-bfloat16-vulnerable}
\end{figure}

To better understand the impact of bit-flip locations, we also evaluate other data types that are frequently used in LLM inference. 
\textbf{Takeaway 15: The ability to recover from error propagation ranks from highest to lowest as quantized integers, \texttt{\small float16}, \texttt{\small float32}, and \texttt{\small bfloat16}.} 
Figure~\ref{fig:illustrating-why-bfloat16-vulnerable} illustrates why \texttt{\small bfloat16} is more susceptible: its larger exponent field amplifies the impact of a single-bit flip, causing exponentially greater deviations that hinder recovery in subsequent computations. 
This observation aligns with prior studies on precision sensitivity in LLM inference~\cite{sun2025demystifying}.

\subsection{Dense or Mixure-of-Expert (MoE) Structures}
\label{sec:dense-or-moe}

We extend our evaluation from dense to MoE architecture. 
Specifically, we adopt \textit{Phi-3.5-MoE-instruct} (denoted as Phi-MoE), a 41.9B-parameter model developed by Microsoft~\cite{Abdin2024Phi3TR} from the same series as \textit{Phi-3.5-mini-instruct} (Phi-Dense). 
Fault injection is performed using \tech on four NVIDIA A100 GPUs. 
We focus on the ARC and BoolQ benchmarks for clarity, as other tasks exhibit consistently.

\begin{table}[h]
\renewcommand{\arraystretch}{0.9}
\centering
\footnotesize
\caption{Evaluating dense and MoE architectures using \textit{Phi-3.5-instruct} series models with ARC and BoolQ tasks.}
\begin{tabular}{l|l|lll|lll} \toprule
{\bf Task} & {\bf Stage} & \multicolumn{3}{c|}{\bf Phi-Dense} & \multicolumn{3}{c}{\bf Phi-MoE} \\
& & \textit{Ori.} & \textit{Com.}  & \textit{Mem.} & \textit{Ori.} & \textit{Com.}  & \textit{Mem.} \\\midrule
\multirow{2}{*}{\bf ARC} & \textit{Prefill} &  \multirow{2}{*}{85.20} & 84.08 & 83.82 & \multirow{2}{*}{90.43} & 89.13 & 89.77  \\
& \textit{First-Token} & & 81.76 & 84.34 &  & 87.84 & 90.01 \\\midrule
\multirow{2}{*}{\bf BoolQ} & \textit{Prefill} &  \multirow{2}{*}{84.00} & 82.71 &  81.67
 & \multirow{2}{*}{85.81} & 83.75 & 84.31 \\
& \textit{First-Token} & & 79.35 &  83.11 &  & 81.94 & 85.02 \\
\bottomrule
\end{tabular}
\label{tab:dense-or-moe}
\end{table}

Table~\ref{tab:dense-or-moe} reports accuracy across different inference stages, excluding the decode phase since it has negligible influence on reasoning tasks (as discussed in Section~\ref{sec:resilience-analysis}). 
\textbf{Takeaway 16: For computation faults, MoE models exhibit resilience similar to dense counterparts, but for memory faults, MoE models are notably more robust.} 
This difference arises because only 20–30\% of experts are activated during inference on ARC and BoolQ, effectively masking faults that occur in inactive parameters. 
Consistent with~\cite{sun2025demystifying}, we also observe that the expert router is highly sensitive to soft errors: even a single error may alter expert selection and cause severe performance degradation. 
However, the router constitutes a tiny fraction of total parameters (<0.1\%), making such faults rare.

\subsection{Multiple Bit-flips Memory Faults}
\label{sec:multiple-bit-flips}

\begin{figure}[ht]
\centering
\includegraphics[width=1.0\columnwidth]{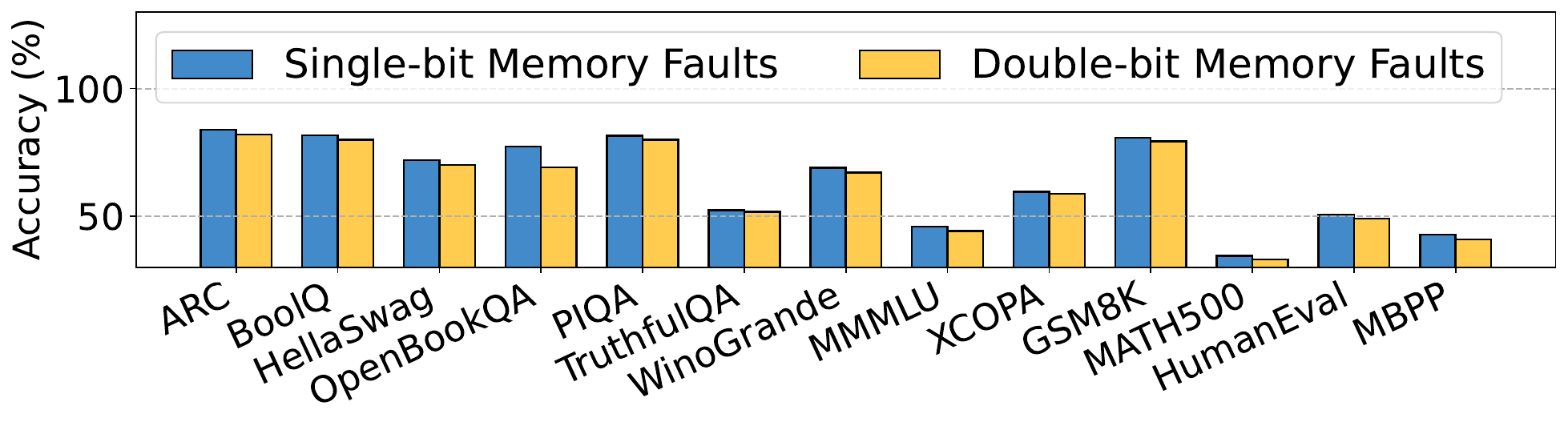}
\vspace{-6mm} 
\caption{Evaluating multiple bit memory faults with Phi.}
\label{fig:multi-bit-memory-results}
\end{figure}

In Section~\ref{sec:fault-model}, we adopt a single bit-flip as the default fault model for both computation and memory faults. 
Although this strategy is widely used for computation faults~\cite{mahmoud2018optimizing,huang2024versatile}, modern memory protection schemes such as ECC can often correct single-bit upsets. 
To better reflect real-world cases, we also evaluate multiple bit-flip memory faults. 
Using Phi over all 13 benchmarks, we inject two bit flips into random indices within the same weight during the prefill stage via \tech. 
Results are shown in Figure~\ref{fig:multi-bit-memory-results}. 
On average, single bit-flip faults yield 63.97 accuracy, while double bit-flip faults slightly reduce this to 61.92, showing consistent trends across all benchmarks. 
After examining layer-wise and bit-location sensitivity, we find that \textbf{Takeaway 17: Multiple bit-flip faults are more likely to corrupt values into irrecoverable numerical patterns, exhibiting slightly lower overall resilience.} 
Note that our multiple bit flips are in the same memory word, typically representing a single parameter (e.g., one \texttt{\small float16}). Bit flips spanning multiple memory words involve more complex fault scenarios and a significantly larger design space, which are beyond the scope of this work. We leave such cases for future study.

\section{Implications for Error Mitigation}

\textit{What practical lessons can we learn from these takeaways?}
Beyond characterizing error propagation behaviors, our results provide concrete guidance for reliable LLM inference.
In this section, inspired by these takeaways, we propose four low-overhead mitigation strategies that require no expensive hardware modifications.

\subsection{Mixed-precision Inference}
In Takeaway~15, we identify that error resilience varies across data types.
This observation naturally raises the question of whether more robust data representations can be leveraged to improve reliability.
However, uniformly adopting such data types may compromise accuracy; for example, quantized integers are often more robust than \texttt{\small float16}, but their reduced numerical precision can degrade accuracy.
Motivated by this, we propose \textbf{Mixed-precision Inference}, of which the key idea is to \textbf{employ different data types for specific components of LLM inference to enhance error resilience while preserving accuracy}.

\begin{figure}[ht]
\centering
\includegraphics[width=1.0\columnwidth]{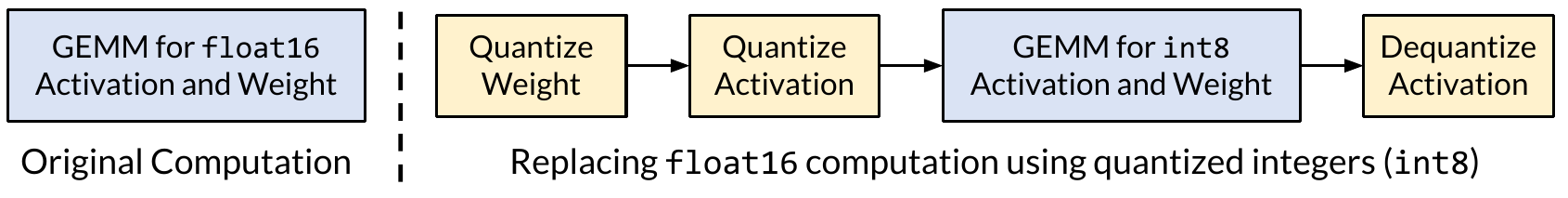}
\vspace{-5mm} 
\caption{Illustrating how to embed quantized integers in original \texttt{\small float16} inference in Mixed-precision Inference.}
\label{fig:mixed-precision}
\end{figure}

Figure~\ref{fig:mixed-precision} illustrates this design.
Compared with the original GEMM using \texttt{\small float16} activations and weights, we first quantize them into \texttt{\small int8} using range-based scaling with fast CUDA kernels.
GEMM is then performed in \texttt{\small int8}, and the resulting integer activations are dequantized back to \texttt{\small float16} for subsequent processing.
To preserve accuracy, this design is selectively applied only to Phi’s \texttt{\small mlp.down\_proj} layers, as motivated by Takeaway~9, while all other layers remain in \texttt{\small float16}.
The \texttt{\small int8} GEMM is implemented using cuBLASLt.
This design naturally extends to other mixed data-type configurations, such as combining \texttt{\small float16} and \texttt{\small float32}.

\begin{table}[h]
\renewcommand{\arraystretch}{0.9}
\centering
\footnotesize
\caption{Evaluating Mixed-precision Inference on ARC and BoolQ for computation faults in the first-token stage.}
\begin{tabular}{l|ll|ll} \toprule
\multirow{2}{*}{\bf Model and Protection} & \multicolumn{2}{c|}{\bf ARC} & \multicolumn{2}{c}{\bf BoolQ} \\
& \textit{w/o Fault} & \textit{w/ Fault}  & \textit{w/o Fault} & \textit{w/ Fault} \\\midrule
Phi (w/o Mixed-precision) & \textbf{85.20} & 81.76 & \textbf{84.00} & 79.35\\
Phi (w\ \ \ \   Mixed-precision) & 84.39 & \textbf{83.01} & 82.93& \textbf{81.74}\\
\bottomrule
\end{tabular}
\label{tab:mixed-precision-results}
\end{table}

Table~\ref{tab:mixed-precision-results} summarizes the evaluation results on ARC and BoolQ using the Phi model.
For each task, we report both fault-free and faulty inference accuracy with and without mixed-precision protection.
Faulty trials correspond to computation faults injected during the first-token stage using \tech.
Similar trends are observed across other evaluated settings.
On ARC, Mixed-precision Inference slightly reduces fault-free accuracy from 85.20\% to 84.39\%, while improving faulty inference accuracy from 81.76\% to 83.01\%.
A consistent trend is observed on BoolQ.
These results indicate that selectively altering numerical precision can improve error resilience with limited impact on fault-free accuracy, demonstrating the potential of mixed-precision designs for reliable LLM inference.
We note that more advanced quantization techniques (e.g., GPTQ~\cite{frantar2022gptq}) are orthogonal to our design and may further reduce accuracy loss; exploring such combinations is beyond the scope of this work.

\subsection{Selective Checksum-based Protection}
Checksum-based methods, commonly used in algorithm-based fault tolerance (ABFT),
have been widely studied to improve error resilience in numerical computations~\cite{huang1984algorithm,chen2013online,zhao2020ft}.
Since these methods are specifically designed for GEMM, they have recently been explored to enhance the reliability of LLM~\cite{liang2025attnchecker,titopoulos2025custom,sun2025ft2,dai2025ft,abftforllm}, where self-attention and MLP extensively rely on GEMM.
Their principle can be explained with a checksum operation.
Given a matrix multiplication $Y = AW$, where $A \in \mathbb{R}^{m \times k}$ and $W \in \mathbb{R}^{k \times n}$,
we compute a row-wise checksum of the input activation as $\mathbf{c}_A = \mathbf{1}_m^{T} A$.
This checksum is propagated through the multiplication as $\mathbf{c}_Y^{\mathrm{exp}} = \mathbf{c}_A W$.
After computing the output $Y$, we obtain the actual checksum $\mathbf{c}_Y^{\mathrm{act}} = \mathbf{1}_m^{T} Y$.
A mismatch between $\mathbf{c}_Y^{\mathrm{act}}$ and $\mathbf{c}_Y^{\mathrm{exp}}$ indicates that the GEMM computation has been affected by soft errors.

However, applying checksum-based protection in a comprehensive manner requires non-trivial engineering effort in practical LLM inference.
First, checksum computation introduces additional memory accesses, while LLM inference is already largely memory-bound~\cite{liu2024cachegen,liu2023scissorhands}.
Second, fully integrating checksum operations into all CUDA kernels across diverse LLM architectures would require substantial manual implementation effort.
Moreover, a full checksum-based approach neglects the inherent self-correction capability of transformer layers, as identified in Takeaway~8.
In this work, we introduce \textbf{Selective Checksum-based Protection}.
Guided by Takeaway~9, we \textbf{selectively deploy checksum-based protection only to GEMM operations that are dimension-preserving or dimension-reducing} (e.g., \texttt{\small self\_attn.o\_proj} and \texttt{\small mlp.down\_proj} layers in Phi),
thereby achieving high error coverage while reducing both offline implementation complexity and online deployment overhead.

\begin{figure}[ht]
\centering
\includegraphics[width=0.95\columnwidth]{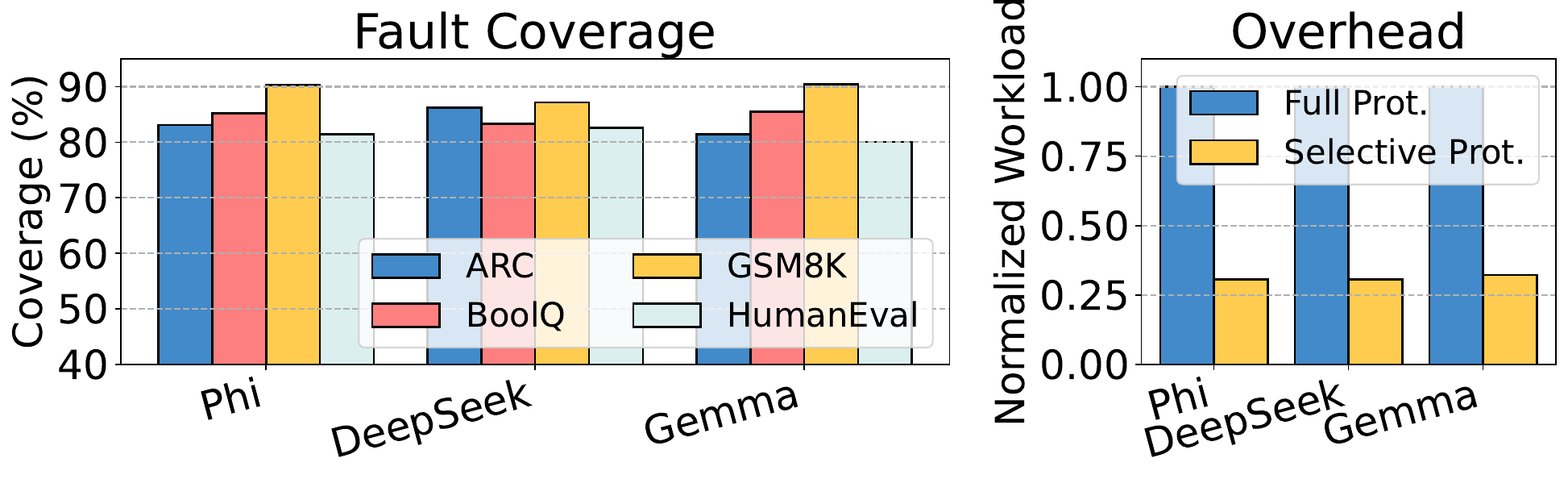}
\vspace{-3mm} 
\caption{Evaluation of fault coverage and overhead of protection with Selective Checksum-based Protection.}
\label{fig:eva-selective-abft}
\end{figure}

Figure~\ref{fig:eva-selective-abft} presents the evaluation results of
Selective Checksum-based Protection.
We evaluate this design from two perspectives: fault coverage and overhead.
Fault coverage is measured as the fraction of fault-induced accuracy degradation
that is mitigated by selectively protecting the identified layers.
We evaluate this metric on ARC, BoolQ, GSM8K, and HumanEval using all three LLMs
studied in this work.
As shown in Figure~\ref{fig:eva-selective-abft} (left), this lightweight strategy
achieves coverage ranging from 80.04\% (HumanEval on Gemma) to 90.31\% (GSM8K on Phi).
To estimate overhead, we compare the GEMM computation workloads of the selected
layers against a hypothetical full checksum-based protection, using FLOPs as a
proxy and normalizing full protection to 1.00.
In Figure~\ref{fig:eva-selective-abft} (right), under this metric, the observed coverage can be achieved while protecting only
30.56\%, 30.57\%, and 32.14\% of the total GEMM workloads for Phi, DeepSeek, and
Gemma, respectively.
These results indicate that selective checksum-based protection can recover most
fault-induced accuracy loss at a fraction of the cost of full protection.

\subsection{Dual First-token Generation}
In Takeaway~3, we observe that the first-token stage in LLM inference is particularly vulnerable. Errors occurring at this stage can significantly alter subsequent decoding behavior, as the first generated token determines the initial trajectory of the autoregressive decoding process. As a result, corruption of the first token can lead to substantial accuracy degradation, reaching up to 4.65\% on Phi for the BoolQ task. 
These observations suggest that improving the reliability of the first-token generation is critical. 
Motivated by this insight and the decoupled inference stages enabled by the design of \tech, we propose \textbf{Dual First-token Generation}. The key idea is to \textbf{reuse language context, generate the first output token twice, and compare two copies before entering the decode loop}, enabling low-cost and reliable detection for LLM inference.

\begin{figure}[ht]
\centering
\includegraphics[width=1.0\columnwidth]{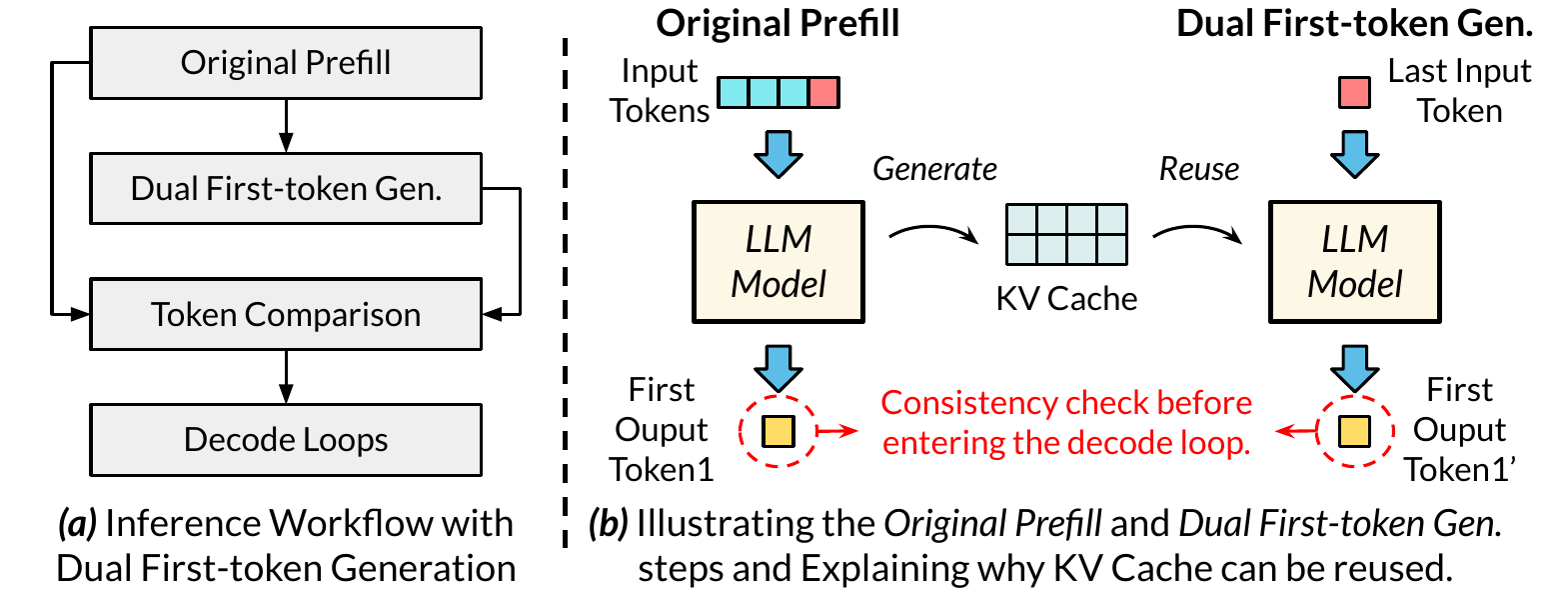}
\vspace{-3mm} 
\caption{Illustrating the Dual First-token Generation.}
\label{fig:dual-first-token-gen}
\end{figure}

Figure~\ref{fig:dual-first-token-gen} illustrates the proposed mitigation strategy.
Given the input tokens, \textit{Original Prefill} processes the entire input sequence, generates the first output token, and stores the language context in the KV cache.
Subsequently, \textit{Dual First-token Generation} reuses the KV cache and takes only the last input token to generate a second copy of the first output token.
Before proceeding to further computation, \textit{Token Comparison} checks whether the two generated tokens are consistent.
If they match, the inference proceeds to the \textit{Decode Loop} to generate subsequent tokens.
Otherwise, a mismatch indicates potential silent corruption caused by soft errors, and the current inference is discarded and retried.
Compared with the original inference workflow, \textit{Dual First-token Generation} and \textit{Token Comparison} are two additional steps.
The former reuses the KV cache and involves only a single-token forward pass through the LLM, while the latter consists of a simple conditional check, resulting in negligible overhead.
Although two copies are sufficient for error detection, recovery can be achieved by extending this mechanism to triple modular redundancy with majority voting~\cite{chang2006automatic}.

\begin{figure}[ht]
\centering
\includegraphics[width=0.95\columnwidth]{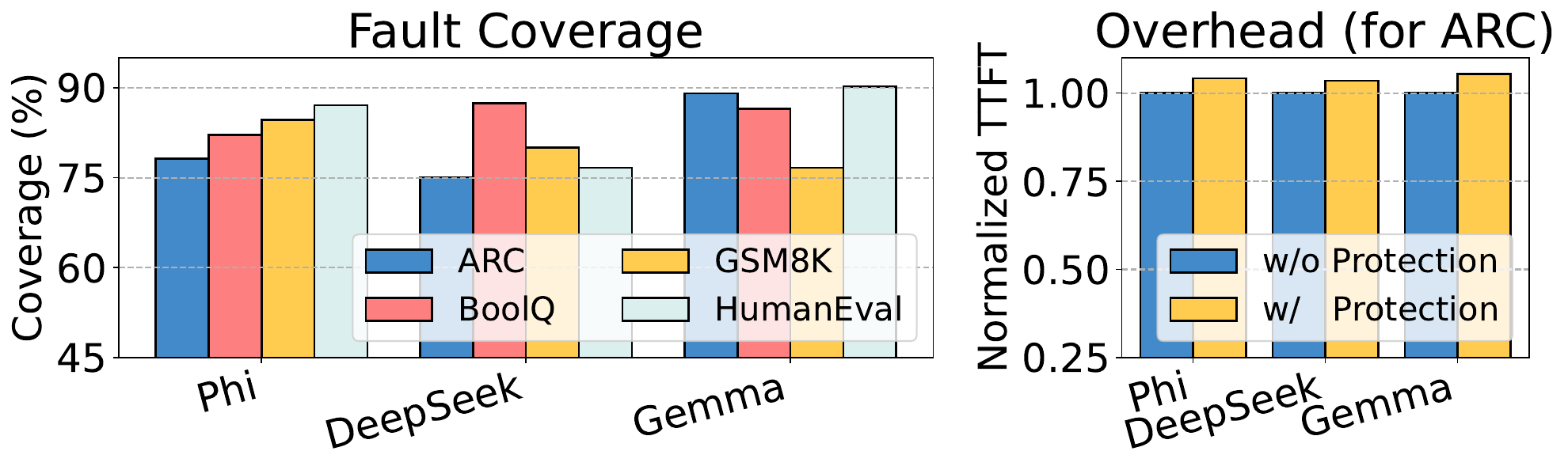}
\vspace{-3mm} 
\caption{Evaluation of fault coverage and overhead of protection with Dual First-token Generation.}
\label{fig:eva-dual-first-token}
\end{figure}

We evaluate this mitigation strategy under computation faults injected during the prefill and first-token stages, with results shown in Figure~\ref{fig:eva-dual-first-token}.
For fault coverage (Figure~\ref{fig:eva-dual-first-token}, left), we measure the fraction of fault-injection trials that lead to accuracy degradation and are successfully identified.
For overhead (Figure~\ref{fig:eva-dual-first-token}, right), we report the Time to First Token (TTFT)~\cite{nvidia-llm-metric}, a critical metric for evaluating LLM inference latency.
As seen, Dual First-token Generation achieves a fault coverage of 82.20\% while incurring only a 3.37\% increase in TTFT, demonstrating an effective trade-off between reliability and performance.
The remaining undetected cases primarily stem from scenarios where faults corrupt the KV cache context during prefill, which can occasionally lead to consistent first-token outputs.
We note that this mechanism is particularly suited for correctness-sensitive inference settings, such as deterministic decoding used in evaluation or safety-critical deployments.


\subsection{Dynamic Shot Inference}

\begin{figure}[ht]
	\centering
	\subfigure[K-shot Example (K=1)]
	{\includegraphics[width=0.32\columnwidth]{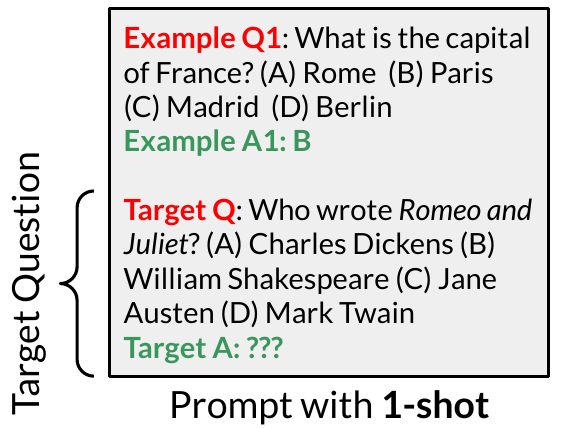}\label{fig:explaining-more-shots}}
	\subfigure[ARC on Phi]
	{\includegraphics[width=0.315\columnwidth]{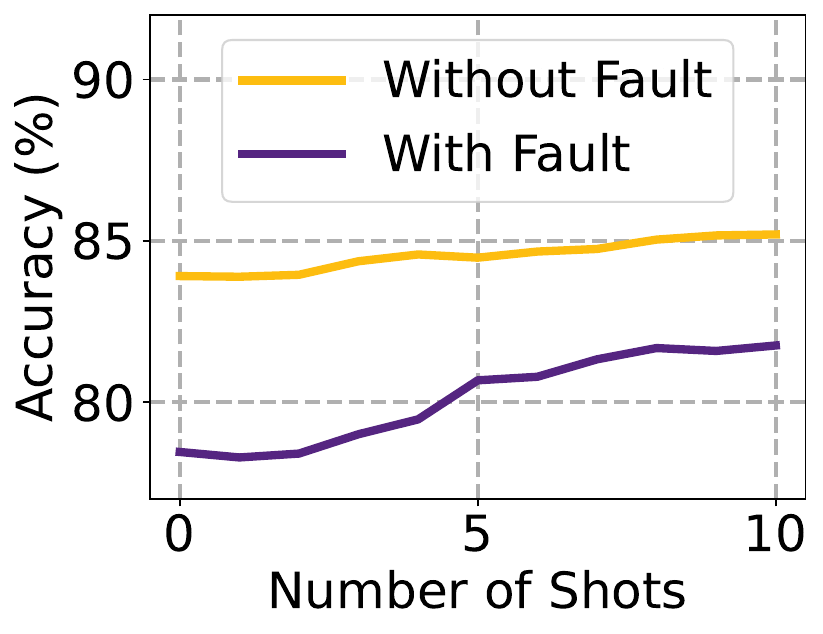}\label{fig:evaluations-arc}}
    \subfigure[BoolQ on Phi]
	{\includegraphics[width=0.315\columnwidth]{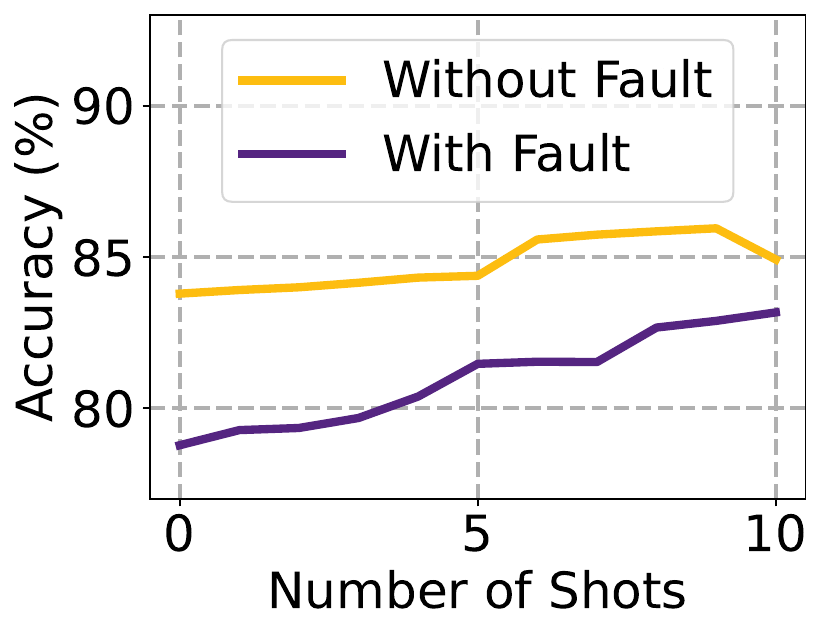}\label{fig:evaluations-boolq}}
    \vspace{-3mm}
	\caption{Illustrating K-shot prompt and evaluating ARC and BoolQ tasks using Phi with different number of shots.}
	\label{fig:dynamic-shot-inference}
\end{figure}

In Table~\ref{tab:resilience-evaluation-table}, we follow the default inference configuration adopted in industry practice, particularly with respect to the number of shots~\cite{Abdin2024Phi3TR}, which leads to the results summarized in Table~\ref{tab:resilience-evaluation-table} and Takeaway~1.
Here, the number of shots refers to the count of in-context examples provided in the prompt before the test query~\cite{brown2020language}, as illustrated in Figure~\ref{fig:explaining-more-shots}.
To examine the impact of shot count on fault resilience, we use \tech to inject computation faults during the first-token generation stage of Phi on the ARC and BoolQ benchmarks.
Our results indicate that increasing the number of shots can mitigate accuracy degradation caused by soft errors.
This observation naturally motivates a lightweight mitigation strategy, termed \textbf{Dynamic Shot Inference}, whose key idea is to \textbf{adaptively increase the number of shots to improve error resilience}.

Figure~\ref{fig:dynamic-shot-inference} presents the evaluation results.
We construct structured prompts with the number of shots ranging from 0 to 10.
On ARC (Figure~\ref{fig:evaluations-arc}), while the fault-free accuracy remains relatively stable, the accuracy under injected faults improves from 78.46\% to 81.76\%.
A similar trend is observed for other multiple-choice tasks (e.g., BoolQ in Figure~\ref{fig:evaluations-boolq}) and generative workloads~\cite{cobbe2021training}.
Increasing the number of shots, however, does not always improve accuracy.
Excessively long prompts may introduce irrelevant context or exceed the model’s effective attention capacity, leading to degraded performance.
These results indicate that, without altering the model architecture or inference mechanism, moderately extending the context length can enhance resilience to soft errors.
We further observe that this effect is more pronounced for faults injected during the prefill stage. Designing an optimal policy for dynamically selecting the number of shots is task dependent and involves additional trade-offs, which we leave for future work.

\section{Conclusion and Future Works}


In this work, we present \tech, a flexible and fine-grained fault injection framework for modeling error propagation in LLM inference.
Through systematic experiments across 13 representative inference tasks and 3 open-weight LLMs, combined with in-depth analysis, we distill 17 takeaways that characterize how computation and memory faults propagate across layers, inference stages, data representations, and inference settings.
Our results reveal pronounced stage- and layer-dependent sensitivity, the corrective and amplifying roles of dimensional transformations, and the strong influence of numerical precision and prompt context on error resilience.
Beyond characterizing these behaviors, our study demonstrates that such empirical insights can directly inform the design of practical mitigation mechanisms.
Guided by the identified takeaways, we introduce four low-overhead, software-level mitigation directions, including Mixed-precision Inference, Selective Checksum-based Protection, Dual First-token Generation, and Dynamic Shot Inference.
These mechanisms selectively target error-sensitive components and stages, achieving improved reliability without requiring hardware modification or comprehensive protection.

Looking forward, our goal is to further explore the vulnerability patterns in LLM inference, such as critical KV cache layers. These insights may inform the design of reliability-aware inference strategies that balance robustness, accuracy, and performance.

\begin{acks}
This work was supported by the U.S. Department of Energy, Office of Science, Advanced Scientific Computing Research (ASCR), under contract DE-AC02-06CH11357 and DE-SC0024559, and by the National Science Foundation under Grants 2540175, 2544839, and 2546265.
The authors also acknowledge the use of the Argonne Leadership Computing Facility (ALCF) Polaris and the Argonne Laboratory Computing Resource Center (LCRC) Swing.
\end{acks}

\bibliographystyle{ACM-Reference-Format}
\bibliography{reference}

\end{document}